\documentclass[Journal,letterpaper]{ascelike-new}
\usepackage[usenames,dvipsnames]{color} %

\usepackage{listings}
\definecolor{code_comment}{rgb}{0,0.2,0.4}
\definecolor{code_string}{rgb}{0.1,0.4,0}
\definecolor{code_standard}{rgb}{0.3,0,0}
 
\usepackage[utf8]{inputenc}
\usepackage[T1]{fontenc}
\usepackage{lmodern}
\usepackage{graphicx}
\usepackage{listings}
\usepackage[figurename=FIG.,labelfont=bf,labelsep=period]{caption}
\usepackage{subcaption}
\usepackage[title]{appendix}
\usepackage{amsmath}
\usepackage{apacite}
\usepackage{url} 
\usepackage{amssymb} 
\usepackage{bbm}
\usepackage{dsfont}
\usepackage{ragged2e}
\usepackage{amsbsy}
\usepackage{newtxtext,newtxmath}
\usepackage{booktabs}
\usepackage[colorlinks=true,citecolor=black,linkcolor=black]{hyperref}
\NameTag{Kley-Holsteg, \today}
\begin{document}

\title{Probabilistic Multi-Step-Ahead Short-Term Water Demand Forecasting with Lasso}
\author[1]{Jens Kley-Holsteg}
\author[2]{Florian Ziel}

\affil[1]{Faculty of Economics, esp. Economics of Renewable Energies, University Duisburg-Essen, Universitätsstr. 2, 45141 Essen, Germany, with corresponding author email. Email: jens.kley-holsteg@stud.uni-due.de}
\affil[2]{Faculty of Economics, esp. Economics of Renewable Energies, University Duisburg-Essen, Universitätsstr. 2, 45141 Essen, Germany.}

\maketitle

\begin{abstract}
Water demand is a highly important variable for operational control and decision making. Hence, the development of accurate forecasts is a valuable field of research to further improve the efficiency of water utilities.
Focusing on probabilistic multi-step-ahead forecasting, a time series model is introduced, to capture typical autoregressive, calendar and seasonal effects, to account for time-varying variance, and to quantify the uncertainty and path-dependency of the water demand process. To deal with the high complexity of the water demand process a high-dimensional feature space is applied, which is efficiently tuned by an automatic shrinkage and selection operator (lasso). It allows to obtain an accurate, simple interpretable and fast computable forecasting model, which is well suited for real-time applications.
The complete probabilistic forecasting framework allows not only for simulating the mean and the marginal properties, but also the correlation structure between hours within the forecasting horizon. For practitioners, complete probabilistic multi-step-ahead forecasts are of considerable relevance as they provide additional information about the expected aggregated or cumulative water demand, so that a statement can be made about the probability with which a water storage capacity can guarantee the supply over a certain period of time. This information allows to better control storage capacities and to better ensure the smooth operation of pumps.    
To appropriately evaluate the forecasting performance of the considered models, the energy score (ES) as a strictly proper multidimensional evaluation criterion, is introduced. The methodology is applied to the hourly water demand data of a German water supplier.
\end{abstract}

\section{Introduction}
\begin{figure}
\centering
\includegraphics[width=0.7\linewidth, angle=270]{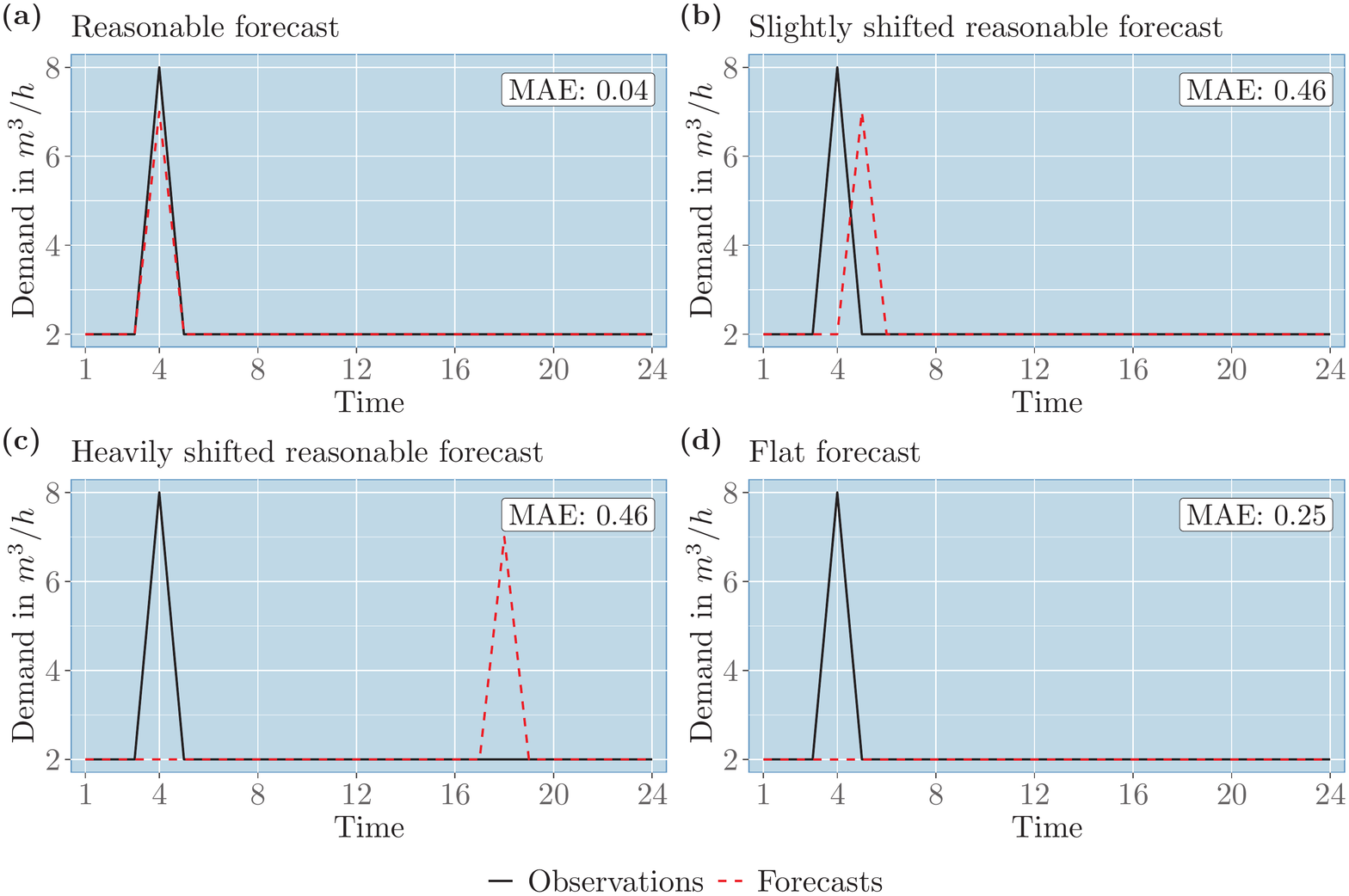}
\caption{Simplified example of water demand forecasts (a)-(d) in the context of water storage management and their evaluated performance in terms of the MAE in $m^3/h$.}
\label{fig_se}
\end{figure}

The increasing availability of high-frequency data in the water sector brings new opportunities to further refine and optimize the efficiency of water utilities. Here, the use of data-driven, short-term water demand forecasting models to reduce energy costs has become a popular field of application in practice, as noted by \citeN{Alvisi.2007}, \citeN{Brentan.2017b}, and \citeN{Arandia.2016}. Reliable information about the expected demand allows, for instance, for optimizing the control of storage capacities to balance demand peaks and to run and schedule pumps more efficiently.\\
The quantification of the underlying uncertainty of future demand forecasts turned out to be of great importance in the decision-making process as emphasized by \citeN{Donkor.2014} and \citeN{Hutton.2015}. In this regard, \citeN{Alvisi.2017} noted that a distinction has to be made between prediction uncertainty and emulation uncertainty. The former denotes the uncertainty associated with the natural variability of the true water demand process and describes the actual quantity of interest for practitioners. The latter denotes the uncertainty which arises and cascades within the data collection and modelling procedure (e.g., measurement/-data uncertainty, parameter uncertainty, or model structure uncertainty), as outlined by \citeN{Hutton.2014}. In contrast to prediction uncertainty, emulation uncertainty must be quantified but marginalized, so that the probabilistic forecaster issues only the expected natural variability of the true water demand process. Here, \citeN{Alvisi.2017}, \citeN{Gagliardi.2017}, and \citeN{Chen.2018} have published promising approaches.\\
However, to date the need for modelling the correlation structure between single hours in a multi-step-ahead forecast has not been addressed. The same applies to the need for appropriate evaluation criteria, which should likewise be able to penalize errors in the mean, the marginal properties and the correlation structure.\\
The practicality of providing a complete probabilistic multi-step-ahead forecast can be illustrated by the planing and management of storage capacities. Here, decision makers are interested in the aggregated demand, so that a statement can be made about the probability with which a water storage capacity can guarantee the supply over a certain period of time. This information forms the foundation to better balance demand peaks and to better schedule the pumping arrangements to take advantage of the electricity price structure. Furthermore, the evaluation of the correlation structure requires more sophisticated evaluation measures. To illustrate the limitations of evaluation measures used so far in the water demand forecasting literature, four hypothetical point forecasts are introduced in Fig. \ref{fig_se}, as done in similar way by \citeN{Haben.2014} in the field of energy demand modelling. As evaluation criterion the well-known mean absolute error (MAE) is applied. In the context of using the forecasts for water storage optimization to balance the expected demand, forecast (a) provides the best fit. Forecast (b) is moderate, and both forecasts on the bottom (c) and (d) are rather poor. Even so, forecast (b) does not exactly hit the peak; it is only slightly shifted, while both forecasts on the right substantially miss the demand peak. However, by considering the MAE, forecast (b), which was assessed as moderate in terms of storage optimization, achieves together with forecast (c) the worst score. Here, the authors introduce, for the following model evaluation, the energy score (ES) as an appropriate evaluation criterion to adequately account for the correlation structure over time, beside the mean and the marginal properties.\\
Focusing on the applied models and methods in the short-term water demand forecasting literature, it is striking that a vast variety of methods and modelling techniques have already been applied to predict future water demand.
Initially linear regression and time series models were used, as outlined by \citeN{Herrera.2010}. However, with advances in the field of machine learning, various types of methods such as artificial neural networks, for example applied by \citeN{Adamowski.2010}, \citeN{Bata.2020}, \citeN{Guo.2018}, \citeN{Ghiassi.2008}, and \citeN{Anele.2017}, support vector regressions used by \citeN{Brentan.2017}, \citeN{Msiza.2008} and \citeN{Shabani.2016}, and random forests, as applied by \citeN{Herrera.2010} and \citeN{Chen.2017}, were successfully introduced. Furthermore, also combinations of the above methods, so-called hybrid-methods have received substantial attention, as applied for example by \citeN{Ambrosio.2019}. Nevertheless, linear regression and time series models are still ranked among the most popular modelling methods, as applied in the recent past by \citeN{Arandia.2016}, \citeN{Chen.2014}, \citeN{Caiado.2010}, and \citeN{Chen.2018}. However, as noted by \citeN{Pacchin.2019} and \citeN{Ghalehkhondabi.2017}, it is still difficult to pick a single method as the overall best, so that the performance of forecasting models based on different forecasting techniques are comparable. \\
Considering the stylized facts of the water demand process, non-stationary and non-linear behavior due to multiple seasonalities, autoregressive and external effects (as for example calendar and weather effects) represent the major challenge for modelling the water demand process and hence, require sophisticated models and methods, as noted by \citeN{Herrera.2010} and \citeN{Adamowski.2012}. Following \citeN{Romano.2014}, in practice parsimonious models, which are able to efficiently adapt to ever-changing operating conditions by applying self-learning ability, are preferred. Hence, in the recent literature, especially non-linear models with a low-dimensional feature space have been applied to model efficiently the complex structure of the water demand process, for example, the artificial neural networks used by \citeN{Pacchin.2019} and \citeN{Alvisi.2017}.\\
In this paper, the authors choose a rather different approach and introduce a high-dimensional feature space in a linear modelling framework. By taking advantage of the least absolute shrinkage and selection operator (lasso), introduced by \citeN{Tibshirani.1996}, the feature space can be automatically tuned, so that an efficient, parsimonious, simple interpretable and fast computable forecasting model is obtained. Hence, the model is well suited for application in real-time operating conditions. Moreover, an appropriate multi-step-ahead forecasting framework, to issue a complete multidimensional probabilistic forecasting distribution, is introduced and appropriate point and probabilistic forecasting measures are used to assess the forecasting performance. In this regard, the strictly proper ES can be highlighted, as it allows for simultaneously penalizing errors in the mean, the marginal properties and the correlation structure of an issued complete probabilistic multi-step-ahead forecast. The paper is organized as follows. Section 2 describes the data and discusses the stylized facts of the water demand process; Section 3 presents the time series model; Section 4 introduces the lasso estimation method; Section 5 outlines the forecasting procedure and presents the applied benchmark models; Section 6 presents appropriate point and probabilistic forecasting evaluation measures and introduces the ES as a strictly proper scoring rule; Section 7 summarizes the results for the calibration and validation period, interprets the proposed model, and discusses the practicality of complete probabilistic multi-step-ahead forecasts; and Section 8 concludes the paper.

\section{Data Description and Stylized Facts}
\label{section:2}

\begin{figure}
\centering
\includegraphics[width=0.7\linewidth, angle=270]{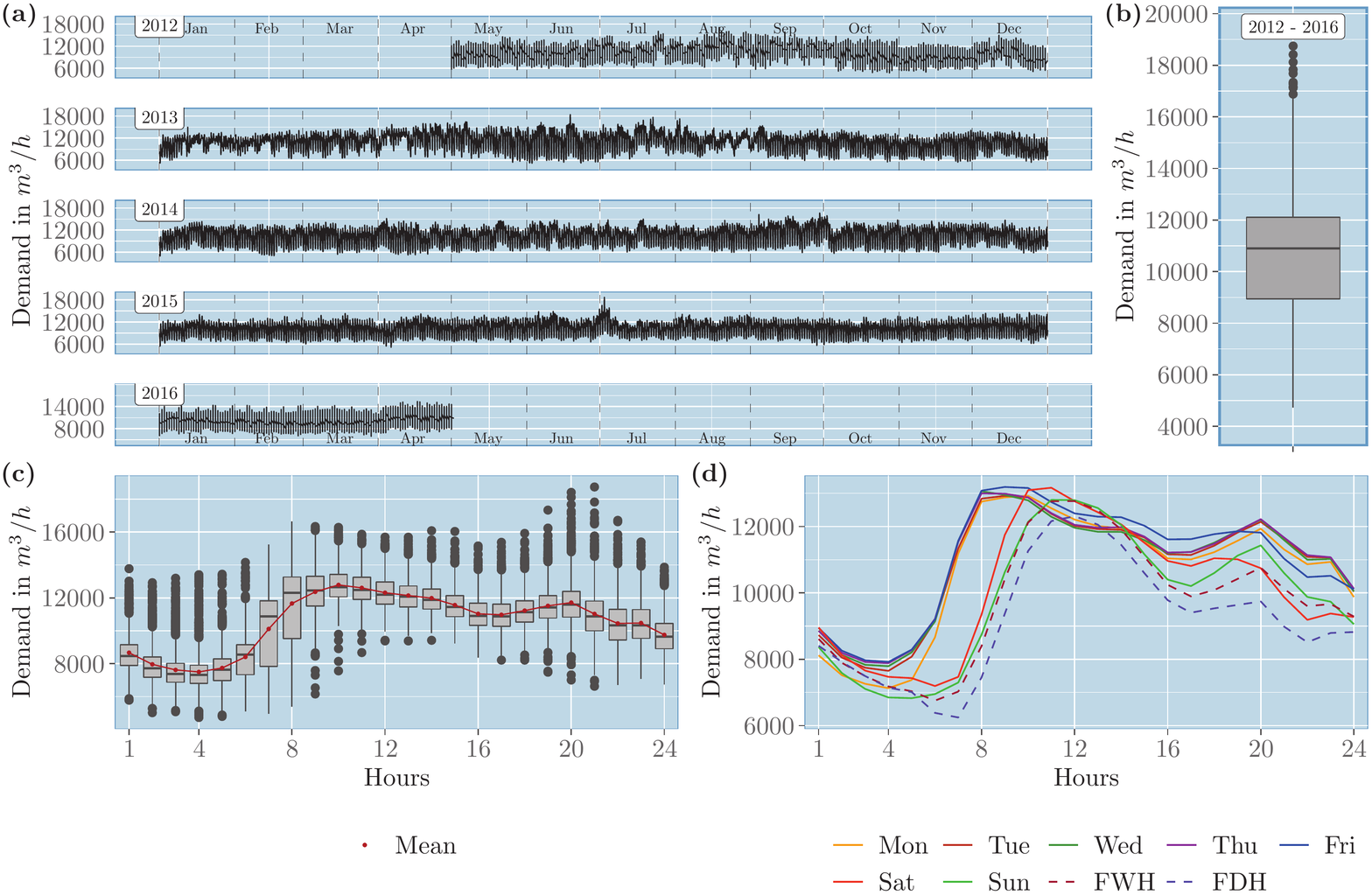}
\caption{Plots of water demand calibration data: time series plot (a), box plot (b), daily periodicity plot (c), and weekly periodicity plot (d).}
\label{fig_timeseries}
\end{figure}

The model is fitted to the hourly water demand data (2012-2018) of a water supplier in the western part of Germany. The data cover six years: four years are used for the model calibration (training) and two years for the model validation (testing). The calibration data are visualized in Fig. \ref{fig_timeseries} (a) and (b). The water distribution system under consideration provides fresh water from one water utility to around one million customers, who use the water for agricultural, private and industrial purposes. The data were not manipulated, so that outliers have not been replaced or transformed. Only the clock change was adjusted to simplify the data analysis.

The water demand process is demand driven and characterized by regional features, which may vary both spatially and temporally from region to region, as outlined by \citeN{Hutton.2015}. As highly influencing factors, the climatic and geographic conditions, as well as the commercial and social conditions of people have been identified, as mentioned by for example \citeN{Anele.2018}. These factors are responsible for the typical observable patterns on a daily, weekly, and an annual basis, as noted by \citeN{Adamowski.2010}. 
Considering the daily structure illustrated in Fig. \ref{fig_timeseries} (c), a trend in the mean but also a varying variance in the day course are observable. The weekly structure illustrated in Fig. \ref{fig_timeseries} (d) is characterized by full working day effects from Tuesday to Thursday and weekend effects on Saturday and Sunday. Friday and Monday are transition days; they are characterized by both, weekend and full working day effects.
Considering the annual patterns, two characteristics are striking. First, the occurrence of holidays has to be highlighted: these days influence the behavior of the demand dramatically, as noted by \citeN{Hutton.2015} and depicted in Fig. \ref{fig_timeseries} (d). The authors classified public holidays in so-called fixed weekday holidays (FWH) and fixed date holidays (FDH), as proposed by \citeN{Ziel.2018}. The former always occurs on the same weekday, but on varying dates, and the latter  always occurs on the same date, but on varying weekdays. Second, meteorological effects might also have a significant influence, as the water demand depends on gradual changes of weather conditions in the year, as illustrated in Fig. \ref{fig_timeseries} (a). Here, a weak seasonal course is observable; especially striking is the high demand in April and the low demand in November. An obvious explanation might be the impact of the growth phase in agriculture on the water demand.

\section{Proposed Model}
To appropriately account for the high-dimensional complexity of the water demand process, the authors propose a model which is characterized by a huge feature space. However, by applying an automatic shrinkage and selection estimation method, the feature space can be tuned, so that a sparse, simple interpretable and fast computable forecasting model is obtained.\\
As inputs, the authors consider autoregressive effects, multiple seasonal patterns and calendar effects. Weather inputs are not considered, as the residual diagnostic indicates no significant effect. This might be explained by the fact that autoregressive effects to some extent already capture the influence of past external inputs such as the gradual increase of temperature over time. However, as indicated by \citeN{Bakker.2014}, an improvement of the forecasting performance by introducing weather inputs is feasible. Therefore, more sophisticated modelling approaches and the inclusion of additional weather forecasts might be required. This, in turn, would imply a significant increase in model complexity and in computational time, which might not be balanced out by the expected benefits, as the proposed model already indicates a good forecasting performance. Hence, the authors refrain in doing so and leave this issue for further research. The proposed model can be defined as

\begin{equation}
Y_t = \mu(t) + \Phi(\mathbb{Y}_t)+ \Psi(t,\mathbb{Y}_t) + \epsilon(t)\label{eq1}
\end{equation}

where $Y_t$ is the water demand at hour $t$ and $\mathbb{Y}_t = (Y_{t-1},Y_{t-2},...) $ denotes the past realizations. The model contains a deterministic component $\mu(t)$, an autoregressive component $\Phi(\mathbb{Y}_t)$, a time-varying autoregressive component $\Psi(t,\mathbb{Y}_t)$ and a zero mean noise process, so that $\mathbb{E}[\epsilon(t)]=0$. The $\epsilon(t)$ component accounts for the stochastic nature of the process, especially for the time-dependent variance structure. To provide a more thorough understanding of the proposed forecasting model, each component of equation (\ref{eq1}) is described in the following subsections.

\subsection{Deterministic component}
\begin{figure}
\centering
\includegraphics[width=1\linewidth]{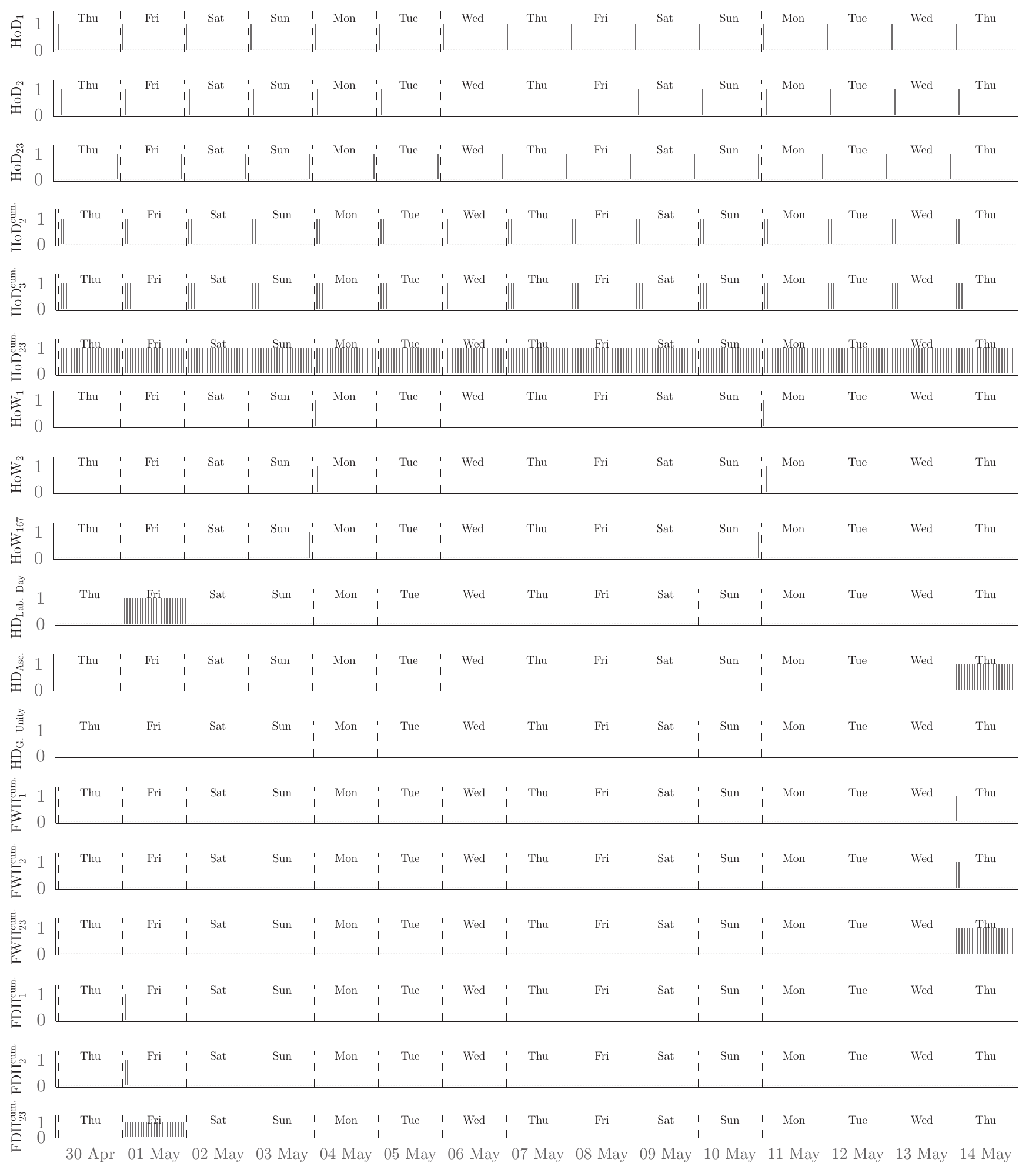}
\caption{Illustration of selected $HoD$-, $HoD^{\text{cum.}}$-, $HoW$-, $HD$-, $FWH^{\text{cum.}}$- and $FDH^{\text{cum.}}$-dummy functions in 2015.}
\label{fig_dummies}
\end{figure}

\begin{figure}
\begin{subfigure}{.5\textwidth}
  \centering
  \includegraphics[width=1\linewidth]{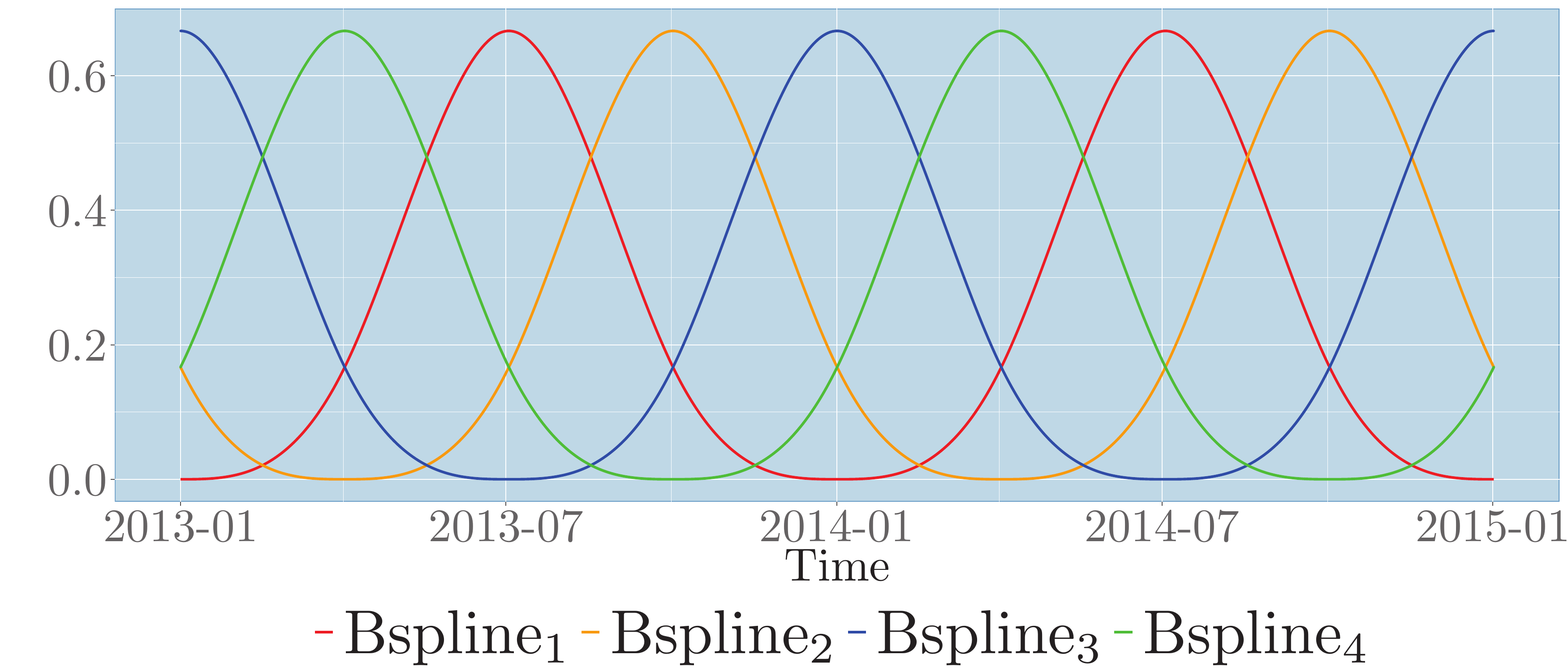}
  \caption{}
  \end{subfigure}%
\begin{subfigure}{.5\textwidth}
  \centering
  \includegraphics[width=1\linewidth]{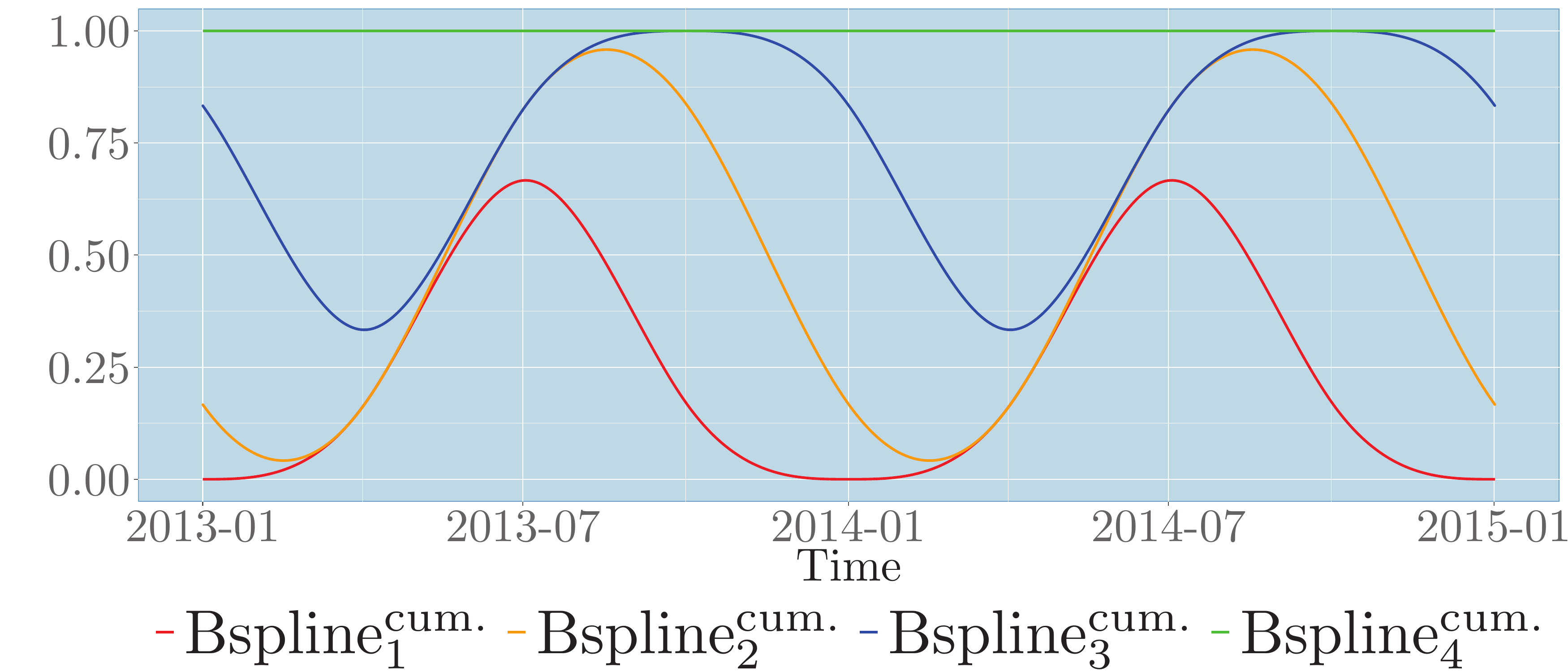}
  \caption{}
  \end{subfigure}
\caption{Cubic B-spline basis (a) and cumulative cubic B-spline basis (b) within a year.}
\label{fig:figbspline}
\end{figure}

As noted in section 2, the water demand process is characterized by strong periodic patterns on a daily, weekly and an annual scale. To control for these non-stationary features, the proposed model includes a deterministic component $\mu(t)$, which varies with time $t$. It can be defined as 
  
\begin{equation}
\begin{split}
\mu(t) & = \mu_0 +\underbrace{\sum_{i=1}^{24}\beta_{i}HoD_{i}(t)}_{\text{\scriptsize Daily effects}}+\underbrace{\sum_{i=1}^{24}\beta_{i+24}HoD_i^{\text{cum.}}(t)}_{\text{\scriptsize Daily effects (cum.)}}+\underbrace{\sum_{i=1}^{168}\beta_{i+48}HoW_{i}(t)}_{\text{\scriptsize Weekly effects}}\\
& + \Pi(t) + \Upsilon(t)\\
\end{split}
\label{mu}
\end{equation}

where $\mu_0$ denotes a constant. The daily and weekly structure is modelled by the hour of the day $HoD$ and the hour of the week $HoW$ dummy functions, respectively. As the lasso approach is especially sensitive to changes over time, cumulative dummies are also introduced. For illustration purposes, the $HoD$, $HoD^{\text{cum.}}$, and $HoW$ dummies are illustrated in Fig. \ref{fig_dummies} for a period of two weeks.\\
The annual structure is modelled by a public holiday component $\Pi(t)$ and a meteorological component $\Upsilon(t)$, which are defined in the two equations below.
The modelling of public holidays poses a major challenge in water demand forecasting, as each holiday is characterized by an individual structure and simultaneously by a rare occurrence. Hence, the authors defined the public holiday component $\Pi(t)$, so that the individual daily structure is reasonably covered, but the risk of over fitting is rather low. The public holiday component is defined as

\begin{equation}
\Pi(t)= \underbrace{\sum_{i=1}^{P}\pi_{i}HD(t)}_{\text{ Holiday effects}} + \underbrace{\sum_{i=1}^{W}\pi_{i+P}FWH_i^{\text{cum.}}(t)}_{\text{\parbox{8em}{\centering\scriptsize Fix weekday holiday effects (cum.)}}} + \underbrace{\sum_{i=1}^{V}\pi_{i+P+W}FDH_i^{\text{cum.}}(t)}_{\text{Fix date holiday effects (cum.)}}\\
\label{PH}
\end{equation}

where $HD(t)$ denotes a dummy function, which models every single public holiday as a function of the day rather than a function of the hour. This allows for the individual characteristic of each day. To adequately capture the hourly characteristics, two additional dummy functions are introduced, namely the $FWH^{\text{cum.}}(t)$ and the $FDH^{\text{cum.}}(t)$ dummy function. The former is the cumulative hour of the day dummy function of the fixed weekday holidays and  the latter of the fixed date holidays. By grouping the holidays into two classes, the authors are able to model the hourly patterns on similar holiday types. To illustrate the public holiday component $\Pi(t)$, each sub-component is visualized for an example period of two weeks in Fig. \ref{fig_dummies}.\\
The meteorological component is modelled by a linear combination of $K$ basis functions, as proposed by \citeN{Ziel.2016b} in the framework of electricity demand forecasting. This component can be defined as 
 
\begin{equation}
\Upsilon(t)=\underbrace{\sum_{i=1}^{K}\upsilon_i B_i^{\text{cum.}}(t)}_{\text{ Annual effects (cum.)}}
\end{equation}

where $\upsilon_{i}$ denotes the parameter for each cumulative basis function $B_i^{\text{cum.}}$. This approach enables the modelling of recurring events in a smooth manner. It increases the flexibility but simultaneously guarantees interpretability of the model. Focusing on the meteorological cycle $K = 4$, cumulative basis functions are used to model each season of the year in a local set-up, as a cold winter, for example, does not necessarily indicate a cold summer, and vice versa. For a detailed explanation of how the basis functions are computed, compare Appendix \ref{appendix:A1} and \citeN{Ziel.2016c}, respectively. To illustrate the considered standard and cumulative basis functions, an example period of one year is visualized in Fig. \ref{fig:figbspline}. Here it can be noted that other approaches such as Fourier series or wavelet compositions might lead to similar results, as applied by \citeN{Alvisi.2007} and \citeN{Adamowski.2012}. 

\subsection{Autoregressive component}
The autoregressive component is the short-term memory of the process and for the proposed model the most valuable component, as highlighted in Fig. \ref{fig:features}, it can be defined as

\begin{equation}
\Phi(\mathbb{Y}_t)=\sum_{k\in \mathbb{I}}\phi_kY_{t-k}
\label{phi}
\end{equation}

where $\phi_k$ is the parameter of the past water demand $Y_{t-k}$ from the index set $\mathbb{I}$. Although the estimation method described below is able to automatically select relevant features from a huge feature space, a pre-selection of index sets is required. The pre-selected lags are presented in Table \ref{lags}.\\ To account for the linear dependency structure of the recent past, the first 361 lags are included. Moreover, seasonal autoregressive components are likewise included to account for the influence of previous weeks and months. 
The pre-selection of lags is based on applying statistical instruments such as the autocorrelation function (ACF) and partial autocorrelation function (PACF).

\subsection{Time-varying autoregressive component}
The time-varying autoregressive component $\Psi(t,\mathbb{Y}_t)$ is in fact an interaction term, containing the most relevant features of the deterministic and autoregressive part and can be defined as

\begin{equation}
\Psi(t,\mathbb{Y}_t) = \sum_{k\in \mathbb{S}}\psi_k(t)Y_{t-k}
\end{equation}
where $\psi_k(t)$ denotes the interaction term of lags of index set $\mathbb{S}$, $\mathbb{S} \subseteq \mathbb{I}$ and the deterministic dummy functions of $HoD(t)$ and $HD(t)$. The selection of the interaction terms was based on the findings of the automatic parameter selection and shrinkage algorithm of the lasso estimator for the deterministic and autoregressive components. 

\begin{table}[htbp]
\caption{Considered lags of index sets $\mathbb{I}$, $\mathbb{K}$, and $\mathbb{S}$}
\centering
\begin{tabular}{l c}
\hline
  Index sets & Lags  \\
\hline
   $\mathbb{I}$  & 1:361,504:505,672:673,840:841,1008:1009,1176:1177,1344:1345\\
   $\mathbb{K}$ & 1:361 \\
   $\mathbb{S}$  & 1,2,24,25,168,169   \\
\hline
\multicolumn{2}{l}{}
\label{lags}
\end{tabular}
\end{table}

\begin{table}[htbp]
\caption{Feature matrix $\boldsymbol{\tilde{X}}_{\text{Cond. mean}}$}
\centering
\scalebox{0.8}{
\begin{tabular}{l l}
\hline
  $\boldsymbol{\tilde{X}}_{\text{Cond. mean}} $ & $= \{1,HoD,HoD^{\text{cum.}},HoW,B^{\text{cum.}},HD,FWH^{\text{cum.}},FDH^{\text{cum.}},Y_{t-\mathbb{I}},Y_{t-\mathbb{S}}HoD, Y_{t-\mathbb{S}}HD \}$ \\
\hline
   $HoD $ & $=\{HoD_1,...,HoD_{23}\}$ \\
   $HoD^{\text{cum.}} $ & $= \{HoD_2^{\text{cum.}},...,HoD_{23}^{\text{cum.}}\}$ \\
   $HoW $ & $= \{HoW_1,...,HoD_{167}\}$ \\
   $B^{\text{cum.}} $ & $= \{B_1^{\text{cum.}},...,B_3^{\text{cum.}}\}$ \\
   $HD $ & $= \{HD_1,...,HD_{11}\}$ \\
   $FWH^{\text{cum.}} $ & $= \{FWH_1^{\text{cum.}},...,FWH_{23}^{\text{cum.}}\}$ \\
   $FDH^{\text{cum.}} $ & $= \{FDH_1^{\text{cum.}},...,FDH_{23}^{\text{cum.}}\}$ \\
   $Y_{t-\mathbb{I}} $ & $= \{{Y_{t-k} | k \in \mathbb{I}}\}$\\
   $Y_{t-\mathbb{S}}HoD$ & $= \{{Y_{t-s} HoD_{k} | s \in \mathbb{S},k \in \{1,...,24}\}$\\
   $Y_{t-\mathbb{S}}HD$ & $= \{{Y_{t-s} HD_{k} | s \in \mathbb{S},k \in \{1,...,11}\}$\\
\hline
\multicolumn{2}{l}{}
\label{xcmean}
\end{tabular}}
\end{table}

\begin{table}[htbp]
\caption{Feature matrix $\boldsymbol{\tilde{X}}_{\text{Cond. variance}}$}
\centering
\scalebox{0.8}{
\begin{tabular}{l l}
\hline
  $\boldsymbol{\tilde{X}}_{\text{Cond. variance}} $ & $= \{1,HoD,HoW,B,HD,FWH,FDH,Y_{t-\mathbb{K}}\}$ \\
\hline
   $HoD $ & $=\{HoD_1,...,HoD_{23}\}$ \\
   $HoW $ & $= \{HoW_1,...,HoD_{167}\}$ \\
   $B $ & $= \{B_1,...,B_3\}$ \\
   $HD $ & $= \{HD_1,...,HD_{11}\}$ \\
   $FWH $ & $= \{FWH_1,...,FWH_{23}\}$ \\
   $FDH $ & $= \{FDH_1,...,FDH_{23}\}$ \\
   $Y_{t-\mathbb{K}}$ & $=\{Y_{t-k}| k \in \mathbb{K}$\}\\
\hline
\multicolumn{2}{l}{}
\label{Xcvar}
\end{tabular}}
\end{table}

\subsection{Time-varying zero mean noise process}
As noted in section \ref{section:2} and depicted in Fig. \ref{fig_timeseries} (c), the water demand process suffers from heteroscedasticity. To account for the conditional time-varying variance, the error term $\epsilon(t)$ is modelled as a function of time, assuming that the variance in time point $t$ depends on the variance of past hours. Here, a time-varying ARCH(p) process is used, as suggested by \citeN{Ziel.2015} in the context of energy market modelling. Then the error term $\epsilon(t)$ can be written as\\

\centerline{
$\epsilon(t) = \sigma(t)Z_t$  where $Z_t$ is i.i.d. with $\mathbb{E}(Z_t)=0$ and $\mathbb{V}$ar $(Z_t)=1.$}  

The conditional variance $\sigma^2(t)$ is modelled as 
\begin{equation}
\sigma^2(t) = \alpha_0(t) + \sum_{k\in \mathbb{K}}\alpha_k \epsilon^2_{t-k}
\label{eq3}
\end{equation}
where $\alpha_k$ is $\geq0$. This is required as the variance can not become negative. 
To model the autoregressive structure, the authors use the lags from index set $\mathbb{K}$. As $\mathbb{K} \subseteq \mathbb{I}$, the lag pre-selection is based on the pre-selection in equation (\ref{phi}). The deterministic part $\alpha_0(t)$ is defined as

\begin{equation}
\begin{split}
\alpha_0(t) &= \theta_0 +\underbrace{\sum_{i=1}^{24}\theta_{i}HoD_{i}(t)}_{\text{\scriptsize Daily effects}}+\underbrace{\sum_{i=1}^{168}\theta_{i+24} HoW_{i}(t)}_{\text{\scriptsize Weekly effects}} +  \Gamma(t)
+ \Xi(t)
\end{split}
\label{a_0}
\end{equation}

where $\alpha_0(t)$ equals the deterministic component of equation (\ref{mu}), adjusted for the cumulative components. This is understandable, as changes over time require the existence of negative parameters, which in turn are ruled out by the non-negativity constraint. Hence, $\Gamma(t)$ is the non-cumulative public holiday component, which is defined as

\begin{equation}
\Gamma(t)= \underbrace{\sum_{i=1}^{P}\gamma_{i}HD(t)}_{\text{Holiday effects}} +\underbrace{ \sum_{i=1}^{W}\gamma_{i+P}FWH_i(t)}_{\text{\parbox{8em}{\centering\scriptsize Fix weekday holiday effects}}} + \underbrace{ \sum_{i=1}^{V}\gamma_{i+P+W}FDH_i(t)}_{\text{Fix date holiday effects}}\\
\label{PH_var}
\end{equation}

and $\Xi(t)$ is the non-cumulative meteorological component, which is defined as

\begin{equation}
\Xi(t)=\underbrace{\sum_{i=1}^{K}\xi_i B_i(t)}_{\text{Annual effects}}.
\label{annual_var}
\end{equation}

\section{Estimation method}
 
As mentioned above, the authors use the lasso estimator and its properties for handling huge feature spaces to obtain a parsimonious and efficient model. The lasso algorithm is able to distinguish between relevant and irrelevant features with the so-called selection ability. Moreover, the lasso algorithm is also able to weight the features in accordance to their explanatory power for the independent variable. This is denoted as the shrinkage ability. Hence, irrelevant features will be fully excluded and less important features lose influence, as outlined by \citeN{Hastie.2015}. The lasso is the standardized version of the well-known ordinary least squares (OLS) estimator, extended by a penalty term. The lasso optimization problem can be defined as

\begin{equation}
\widehat{\boldsymbol{\tilde{\beta}}}_{\lambda}^{lasso} = \operatorname*{arg\ min}_{\beta}||\boldsymbol{\tilde{Y}}-\boldsymbol{\beta}^{\intercal}\boldsymbol{\tilde{X}}||_2^2 + \lambda||\boldsymbol{\beta}||_1
\end{equation}

where $\boldsymbol{\tilde{Y}}$ denotes the standardized water demand vector, $\boldsymbol{\tilde{X}}$ the standardized feature matrix and $\widehat{\boldsymbol{\tilde{\beta}}}$ the estimated standardized lasso parameter vector. For illustration purposes, in Table \ref{xcmean} the feature matrix for the conditional mean estimation and in Table \ref{Xcvar} the feature matrix for the conditional variance estimation are depicted. The tuning parameter $\lambda$ regulates the impact of the penalty term. In case $\lambda$ converges to zero, the penalty term becomes meaningless and the standard OLS solution is obtained. However, as $\lambda$ approaches one, the penalty term forces the estimation parameters one by one to become exactly zero. This solution would imply that no parameters are included in the model. To choose the appropriate tuning parameter $\lambda$, a selection algorithm is required. Here, the authors use the Bayesian information criterion (BIC), which is especially suitable for huge feature spaces as it is more conservative than, for example, the well-known Akaike information criterion (AIC). This is a valuable property, as especially the risk of over fitting increases with a high model parametrization. 
For further details about the BIC, see \citeN{Neath.2012}.

For computation, the authors apply the glmnet package in R. It is based on the coordinate descent algorithm, which enables a fast and efficient parameter estimation. For further explanation, see \citeN{Friedman.2010} and \citeN{Hastie.2015}. 

\section{Forecasting Set-up}
As noted by \citeN{Alvisi.2017} the quantity of interest in probabilistic forecasting is the natural variability of the true water demand process and not the uncertainty arising and cascading in the modelling procedure. Based on this, complete probabilistic multi-step-ahead forecasts are constructed, which consider beside the mean and the marginal properties, especially the correlation structure between each issued hour $h \in H$, so that a joint distribution rather than a marginal distribution is issued. This implies that the expected water demand for a fixed forecasting horizon $H>1$, can be seen as a multivariate random variable $\boldsymbol{Y} \in R^H$, where $\boldsymbol{Y}$ follows an unobservable distribution $G$, so that $\boldsymbol{Y} \sim G$.\\ 
To appropriately model the multivariate random variable $\boldsymbol{Y} \in R^H$, ensemble forecasts reveal preferable properties. In the water demand forecasting literature, ensemble forecasts are commonly used, for example by \citeN{Tiwari.2013} and \citeN{Hutton.2014}. However, these forecasts have predominately been applied to account for uncertainties arising in the emulation process.\\
In this research study, ensemble forecasts are used to model the prediction uncertainty. As forecasting horizon, $H=24$ is chosen and an ensemble is created by recursively solving the corresponding forecasting model in a Monte-Carlo simulation with a total of $M=1,000$ sample paths. To illustrate the meaning of the correlation structure, different model simulations are depicted in the left column of Fig. \ref{fig_CD}. In addition to the standard procedure (a), three manipulated correlation structures are introduced. The comonotone model simulation is characterized by a perfect positive path-dependency (b), so that the sample paths are non-intersecting, that is, the path that is highest in the first hour remains highest for the following hours and so on. The countermonotone model simulation (c), by comparison, considers perfect negative pairwise dependency, so that at each hour the sample paths run contrary to the hour before, that is, the path that is highest in the first hour is lowest in the second and so on. Finally, the independent model simulation (d) is characterized by an independent path-dependency, so that each hour of a sample path is arbitrarily connected to the adjacent hours.\\

The forecasting study is based on $1,000$ equally distributed forecasting time points drawn from the validation data to mimic real-time conditions. To account for ever changing operating conditions a rolling window approach is used to ensure a recurring parameter re-estimation at every forecasting time point, as suggested e.g. by \citeN{Pacchin.2017}. Here, it might be worth noting that the forecasting time point is not starting every corresponding day at midnight but instead on varying hours over the day.

\subsection{Benchmarks}
As benchmarks, different forecasting models for hourly water demand data from the literature are applied. First, two SARIMA(0,1,4)(0,1,1) models with a seasonal period of 24 and 168 are considered, based on \citeN{Arandia.2016}. As the computational time with a calibration data length of up to four years distinctly exceeds the requirements for real-time applications, the authors decided to shorten the length of the calibration data for all considered competitors from the literature. Here, \citeN{Arandia.2016} have suggested various data lengths; for the corresponding data a length of 28 days reveals the best results. The time series models are estimated and predicted by applying the forecast package in R, for further explanations see \citeN{Hyndman.2008} and \citeN{Hyndman.2019}.\\
Second, three typical machine learning methods are used, namely an artificial neural network $ANN_{Her}$, a support vector machine $SVM_{Her}$ and a random forest $RF_{Her}$. All three models are based on \citeN{Herrera.2010}. As features, the authors use the lags $1,2,24,168$ and a Fourier series of order $1$ for the daily and annual cycles, respectively. The model tuning parameters and the applied length of the calibration data is chosen in accordance with \citeN{Herrera.2010}, whereby the best model is chosen ex post. The support vector regression is estimated by applying the e1071 package, the neural network is estimated by applying the nnet package, and the random forest is estimated by applying the svm package in R, for further explanations see \citeN{Meyer.2019}, \citeN{Venables.2002}, and \citeN{Liaw.2002}, respectively.\\
Third, a neural network $ANN_{Pac}$ based on \citeN{Pacchin.2019} and \citeN{Alvisi.2017} is applied. A log sigmoid transfer function is used in the hidden layer and a pure linear one in the output layer. The length of the calibration period and the features are chosen based on \citeN{Pacchin.2019}. The number of hidden neurons is set equal to 72 in accordance to \citeN{Alvisi.2017} and the learning rate varied between 0.0001, 0.001, 0.01, and 0.1. Here, the best results are obtained expost for a learning rate of 0.01. The neural network is also estimated by applying the nnet package in R.\\
Fourth, three naive models are applied, namely a naive mean model $Naive_{Mean}$ as used by \citeN{Pacchin.2019} and \citeN{Gelazanskas.2015}, a naive mean model conditioned on the type of the day $Naive_{FM}$, and a naive mixed random walk model $Naive_{MRW}$. The $Naive_{Mean}$ model computes the values of the forecast by taking the arithmetic average of each hour of the day. The $Naive_{FM}$ computes likewise the $Naive_{Mean}$ model the forecast based on the arithmetic average of each hour of the day, however, with distinction in the type of the day. Here, the authors have distinguished in Mondays, Tuesdays to Thursdays, Fridays, Saturdays, Sundays, and Holidays. The $Naive_{MRW}$ is the only non-parametric model and is defined as

\begin{equation}
Y_t = \begin{cases} Y_{t-168} + \epsilon_t, & \text{on Monday, Saturday or Sunday}\\
                    Y_{t-24} + \epsilon_t, & \text{otherwise}
       \end{cases},
\end{equation}

where $\epsilon_t$ denotes the error term. Finally, also two $AR(p)$ processes are applied. As demonstrated by \citeN{Ziel.2015} in the field of electricity price forecasting, such models are easy to implement and their predictive accuracy is reasonable good. To provide an easily implementable and fast computable forecasting benchmark model in the field of short-term water demand forecasting, the authors have attached the corresponding implementation in R in Appendix \ref{appendix:B}.\\
The $AR(p)$ benchmark models differ in the mean adjustment, so one part includes a mean adjusted hour of the day $D$ process and the other a mean adjusted hour of the week $W$ process. The considered $AR(p)$ processes are estimated by the Yule-Walker estimator and computed by applying the stats package in R, for further explanation see \citeN{R.2019}. The $AR(p)$ models are given by

\begin{equation}
Y_t = \mu^{i}_t  + \sum_{k=1}^{p}(\phi_k^i Y_{t-k}-\mu^{i}_t) + \epsilon_t^{i}
\end{equation}
where $i \in \{D,W\}$. The autoregressive structure is captured by including lags of order $p$, which in turn are chosen by applying the BIC. Here, the maximum order $p$ is set equal to 1,500. All introduced benchmark models are based on a bootstrap distribution assumption.

\Section{Evaluation criteria}
As the evaluation procedure is of particular relevance to identify the best forecasting model, the choice of an appropriate criteria should always be done in accordance with the forecasting purpose at hand. In the framework of point forecasting evaluation, a broad variety of measures is available. Here, the water demand forecasting community is mainly focused on measures such as the MAE, the mean squared error, the Nash-Sutcliffe model efficiency coefficient (NS) and generalizations of the previously mentioned measures as applied by, for example \citeN{Anele.2018}, \citeN{Donkor.2014}, \citeN{Herrera.2010}, and \citeN{Brentan.2017}. Here, it might be worth noting that although the issued forecasts have a probabilistic distribution, they can easily be reduced to simple point forecasts. Here, the MAE and the root mean squared error (RMSE) can be highlighted, as they reveal preferred properties, as discussed by \citeN{Franses.2016}. The MAE is a strictly proper criterion for the median and the RMSE is a strictly proper criterion for the mean. In this context, "strictly proper" refers to the ability that only the perfect forecast minimizes the named criterion, as discussed by \citeN{Gneiting.2014}.\\ 
As noted in the introduction, probabilistic forecasting is gaining acceptance. Hence, probabilistic evaluation measures are also required. As pointed out by \citeN{Gneiting.2007}, the aim of probabilistic forecasting is to maximize the sharpness of probabilistic forecasts subject to calibration. Where, calibration denotes the statistical consistency between the issued distribution and the events that materialize and sharpness refers to the concentration of the predictive distribution.
To adequately assess the properties of a probabilistic forecast, strictly proper scoring rules are well suited. They assign a numerical score based on the predictive distribution and the events that materialize to the corresponding forecast.
Here, the authors introduce the pinball score (PB) as an appropriate scoring rule to evaluate the marginal distribution of an issued forecast. Moreover, as illustrated in Fig. \ref{fig_se}, the dependency structure is likewise relevant, especially if the forecast is to be used in the field of storage capacity optimization. Here, the authors introduce the ES as an appropriate scoring rule. 

\subsection{Point forecasting evaluation criteria}
As point forecasting measures the MAE, RMSE, and the NS are applied. They can be defined as

\begin{equation}
MAE = \frac{1}{N}\sum_{j=1}^N MAE_{t_j}, \text{ \ \ \ with\ \ \ } MAE_{t_j}= \frac{1}{H} \sum_{h=1}^H |Y_{t_j+h}-\hat{Y}_{t_j+h}|,   
\label{MAE}
\end{equation}

\begin{equation}
RMSE = \frac{1}{N}\sum_{j=1}^N RMSE_{t_j},\text{ \ \ \ with\ \ \ } RMSE_{t_j} = \sqrt{\frac{1}{H}\sum_{h=1}^H (Y_{t_j+h}-\hat{Y}_{t_j+h})^2 },
\label{RMSE}
\end{equation}

\begin{equation}
NS = \frac{1}{N} \sum_{j=1}^N NS_{t_j}, \text{ \ \ \ with\ \ \ } NS_{t_j} = 1- \frac{\sum_{h=1}^H (Y_{t_j+h}-\hat{Y}_{t_j+h})^2 } {\sum_{h=1}^H (Y_{t_j+h}-\bar{Y}_{t_j})^2},
\label{NS}
\end{equation}

where $Y_{t_j+h}$ denotes the true water demand at forecasting time point $t_j$ and hour $h$, $\bar{Y}_{t_j}$ the mean of the true demand vector $Y_{t_j} = Y_{t_j+h},...,Y_{t_j+H}$, and $\hat{Y}_{t_j+h}$ the estimated water demand at forecasting time point $t_j$ and hour $h$, $H$ is the number of hours issued in a forecast and $N$ the number of forecasting time points $t_j$ in the forecasting study. These named measures are suited to point forecasting, but are not appropriate in terms of probabilistic forecasting, so that more sophisticated criteria are required.

\subsection{Probabilistic forecasting evaluation criteria}
A promising scoring rule, which appropriately considers the marginal properties of probabilistic forecasts is the PB. It is defined as

\begin{equation}
PB = \frac{1}{N}\sum_{j=1}^N PB_{t_j}, \text{ \ \ \ with\ \ \ } PB_{t_j} (\boldsymbol{\tau}) = \frac{1}{H}\frac{1}{L}\sum_{h=1}^H \sum_{i=1}^L(Y_{t_j+h}-\hat{q}_{t_j+h,\tau_i})(\tau_i-\mathbbm{1}_{\{Y_{t_j+h}-\hat{q}_{t_j+h,\tau_i}<0\}}),
\label{PB}
\end{equation} 

where $\hat{q}_{t_j+h,\tau}$ is the issued quantile and $\tau$ the corresponding quantile level. $Y_{t_j}$ is the true water demand at forecasting time point $t_j$, $L$ denotes the number of quantile levels in the dense grid $\boldsymbol{\tau}$, $N$ is the number of forecasting time points in the forecasting study, $H$ is the number of hours issued in a forecast and $\mathbbm{1}$ is an indicator function. Note that the PB converges to the continuous ranked probability score (CRPS) for an infinitely dense and equidistant grid $\boldsymbol{\tau}$. The CRPS in turn is a widely-accepted strictly proper scoring rule for the full distribution function and a generalization of the well-known MAE, as introduced in equation (\ref{MAE}). Hence, for the special case $\tau=0.5$, the PB is a scaled version of the MAE. For further information on the PB, see \citeN{Nowotarski.2018} and \citeN{Steinwart.2011}.\\ 
However, beside the marginal properties in a multi-step-ahead forecasting framework, also the correlation structure between single hours in a corresponding forecasting horizon is highly relevant, as already noted. In this regard, the PB has limitations, as the evaluations of the issued quantiles of each hour are assessed separately. Currently, only a few scoring rules are known, which are also able to account for the path-dependency between hours. In this regard promising candidates are the logarithmic score, the David-Sebastiani score, the variogram score, and the ES, as highlighted by \citeN{Scheuerer.2015} and \citeN{Gneiting.2007}. Focusing on the forecasting purpose of this research, the ES has preferable properties, as it is a strictly proper scoring rule for probabilistic multi-step-ahead forecasts and it is well suited to deal with the issued sample paths of the ensemble forecast. The ES can be defined as 
\begin{equation}
ES (F,\boldsymbol{\mathrm{y}}) = \mathop{\mathbb{E}_F} [||\boldsymbol{\mathrm{P}}-\boldsymbol{\mathrm{y}}||_2]-\frac{1}{2}\mathop{\mathbb{E}_F} [||\boldsymbol{\mathrm{P}}-\boldsymbol{\mathrm{P}}^{\prime}||_2], 
\end{equation}

whereby $F$ denotes the issued multivariate distribution of the forecaster, $\boldsymbol{\mathrm{P}}$ and $\boldsymbol{\mathrm{P}}^{\prime}$ are independent random draws of $F$, $\boldsymbol{\mathrm{y}}$ is the considered observation vector of the considered process, and $\|\cdot\|_2$ denotes the Euclidean norm. The ES is like the PB, a generalization of the CRPS and equals the CRPS in the case of $H=1$. For detailed explanations, see \citeN{Gneiting.2007}, \citeN{Gneiting.2008}, and \citeN{Scheuerer.2015}. In this setting, the ES is estimated by

\begin{align}
\label{eqn:ES}
\begin{split}
\begin{gathered}
\widehat{ES} = \frac{1}{N} \sum_{j=1}^N ES_{t_j}, \text{ \ \ \ with\ \ \ }
\\
\widehat{ES}_{t_j} (F_{t_j},\mathbf{Y}_{t_j}) = \frac{1}{M} \sum_{i=1}^M ||\mathbf{\widehat{Y}}_{t_j,H}^ {[i]} -\mathbf{Y}_{t_j,H} ||_2-\frac{1}{2M^2}\sum_{i=1}^M\sum_{l=1}^M||\mathbf{\widehat{Y}}_{t_j,H}^ {[i]}-\mathbf{\widehat{Y}}_{t_j,H}^ {[l],\prime}||_2,
\end{gathered}
\end{split}
\end{align}

where $\mathbf{Y}_{t_j,H} = (Y_{t_j +1},...,Y_{t_j+H})$ denotes the water demand vector at the forecasting time point $t_j$. $\mathbf{\widehat{Y}}_{t_j,H}^{[i]}$ and $\mathbf{\widehat{Y}}_{t_j,H}^{[l],\prime}$ denote the corresponding $i$-th and $l$-th simulation vector of the simulation matrices $\mathbf{\widehat{Y}}_{t_j,H}$ and $\mathbf{\widehat{Y}}_{t_j,H}^{\prime}$, respectively. $M$ denotes the number of simulations, $H$ the number of hours issued in a forecast and $N$ the number of forecasting time points in the forecasting study.\\
In the following the ES is applied as the determining measure. However, also the PB, the MAE, RMSE, and NS are listed. Although the ES is able to discriminate errors in the mean, the marginal properties and the correlation structure simultaneously, pinpointing specific causes of a poor performance is rather difficult. Here, the PB, the MAE, and the RMSE might provide additional information in terms of diagnostic checking. 

\subsection{Significance test}
To identify differences in the forecasting performance of competitors in a statistically reliable way, the Diebold Mariano (DM) test is applied. As noted by \citeN{Nowotarski.2018}, the DM test is more popular in the framework of point forecasting, but is also applicable to probabilistic forecasting. The DM test is simply an asymptotic $z$-test with the hypothesis that the mean of the loss differential series is zero. On the assumption that the loss differential is covariance stationary, the DM test is asymptotically standard normal.\\
The authors calculated two one-sided DM tests with a significance level of 5 $\%$, to test for significant differences from zero in both directions. For further details of the DM test, see \citeN{Diebold.1995}. 

\section{Results and discussion of practicality of complete probabilistic multi-step-ahead forecasts}
In this section the forecasting performance of the proposed forecasting model $ARXARCHX_{lasso}$ from equation (\ref{eq1}) is presented in comparison to the introduced benchmark models for the calibration and validation data. The results are presented in Table \ref{C_Results} and \ref{V_Results}. As benchmark model  the $AR(p)^W$ model is chosen to compute the improvements in $\%$. Moreover, the diagnostic checking of the $ARXARCHX_{lasso}$ is shown for the calibration period to pinpoint further improvements and to highlight the most influential parameters. Finally, in the discussion section the practicality of complete probabilistic multi-step-ahead forecasts is outlined and illustrated by a storage management problem. 

\begin{table}[ht]
\caption{Forecasting results, considered data length, considered number of parameters, and performance improvements (Imp.) in \% relative to the $AR(p)^W$ model for each forecasting model within the calibration period.}
\centering
\scalebox{0.7}{
\begin{tabular}{lrrrrrrrrrr}
 \hline
 Models & Data length in & Parameters & Parameters & MAE$^*$ in & Imp. in & RMSE$^*$ in & Imp. in & NS$^*$ in & Imp. in \\ 
  & $h$ & (active) & (possible) & $m^3/h$ & \% & $m^3/h$ & \% & \% & \%\\ 
  \hline
$Naive_{FM}$ & 35064 & 24 & 24 & 888.06 & -293.12 & 1175.64 & -269.25 & 68.10 & -30.27 \\ 
  $Naive_{Mean}$ & 35064 & 144 & 144 & 731.92 & -224.00 & 967.78 & -203.96 & 78.35 & -19.78 \\ 
  $Naive_{MRW}$ & 35064 & 0 & 0 & 644.54 & -185.32 & 947.52 & -197.60 & 79.29 & -18.81 \\ 
  $ANN_{Pac}$ & 8760 & 3624 & 3624 & 474.26 & -109.94 & 665.60 & -109.06 & 89.41 & -8.45 \\ 
  $RF_{Her}$ & 1344 & 782897 & >782897 & 294.67 & -30.44 & 405.83 & -27.46 & 96.11 & -1.59 \\ 
  $SARIMA(0,1,4)(0,1,1)_{24}$ & 672 & 6 & 6 & 288.79 & -27.84 & 404.71 & -27.11 & 96.11 & -1.59 \\ 
  $SVM_{Her}$ & 1344 & 630 & >630 & 278.16 & -23.13 & 373.87 & -17.43 & 96.69 & -0.99 \\ 
  $ANN_{Her}$ & 1344 & 71 & 71 & 262.24 & -16.08 & 344.06 & -8.07 & 97.20 & -0.47 \\ 
  $AR(p)^D$ & 35064 & 1280 & 1668 & 230.60 & -2.08 & 320.02 & -0.51 & 97.63 & -0.03 \\ 
  $AR(p)^W$ & 35064 & 401 & 1668 & 225.90 & 0.00 & 318.39 & 0.00 & 97.66 & 0.00 \\ 
  $ARXARCHX_{lasso}$ & 35064 & 200 & 1468 & 216.13 & 4.33 & 300.16 & 5.73 & 97.91 & 0.26 \\ 
  $SARIMA(0,1,4)(0,1,1)_{168}$ & 672 & 6 & 6 & $\textbf{ 185.20 }$ & $\textbf{ 18.02 }$ & $\textbf{ 291.82 }$ & $\textbf{ 8.34 }$ & $\textbf{ 97.93 }$ & $\textbf{ 0.28 }$ \\ 
  \hline
 \multicolumn9l{$^*$ Within the calibration period the forecasting horizon $H$ in equation (\ref{MAE}) (MAE), (\ref{RMSE}) (RMSE), and (\ref{NS}) (NS) is set equal to 1.}\\
\label{C_Results}
\end{tabular}}
\end{table}

\subsection{Results of calibration period}
By examining the forecasting results in terms of the MAE, RMSE, and NS for the calibration period in Table \ref{C_Results}, the forecasting results of the $SARIMA(0,1,4)(0,1,1)$ models are striking. Here, the low scores in terms of the MAE and RMSE and the high scores in terms of the NS are suspicious and suggest model over fitting. This is most likely caused by the fact that a too short calibration period was chosen. Hence, the extension of the calibration period for the named models can be recommended. Focusing on the remaining forecasting models, the proposed $ARXARCHX_{lasso}$ is ranked the best, followed by the proposed simple $AR(p)$ models. The $Naive$ models are ranked the worst. The rather poor performance of the $ANN_{Pac}$ in comparison to the other sophisticated models is caused by the chosen 24-dimensional output layer of the network, which results in a 24-dimensional forecast in the calibration period instead of a one-step-ahead forecast as issued by the remaining sophisticated models. This allows the $ANN_{Pac}$ to better approximate the forecasting error of the validation period.

\begin{figure}
\centering
\includegraphics[width=0.7\linewidth, angle =270 ]{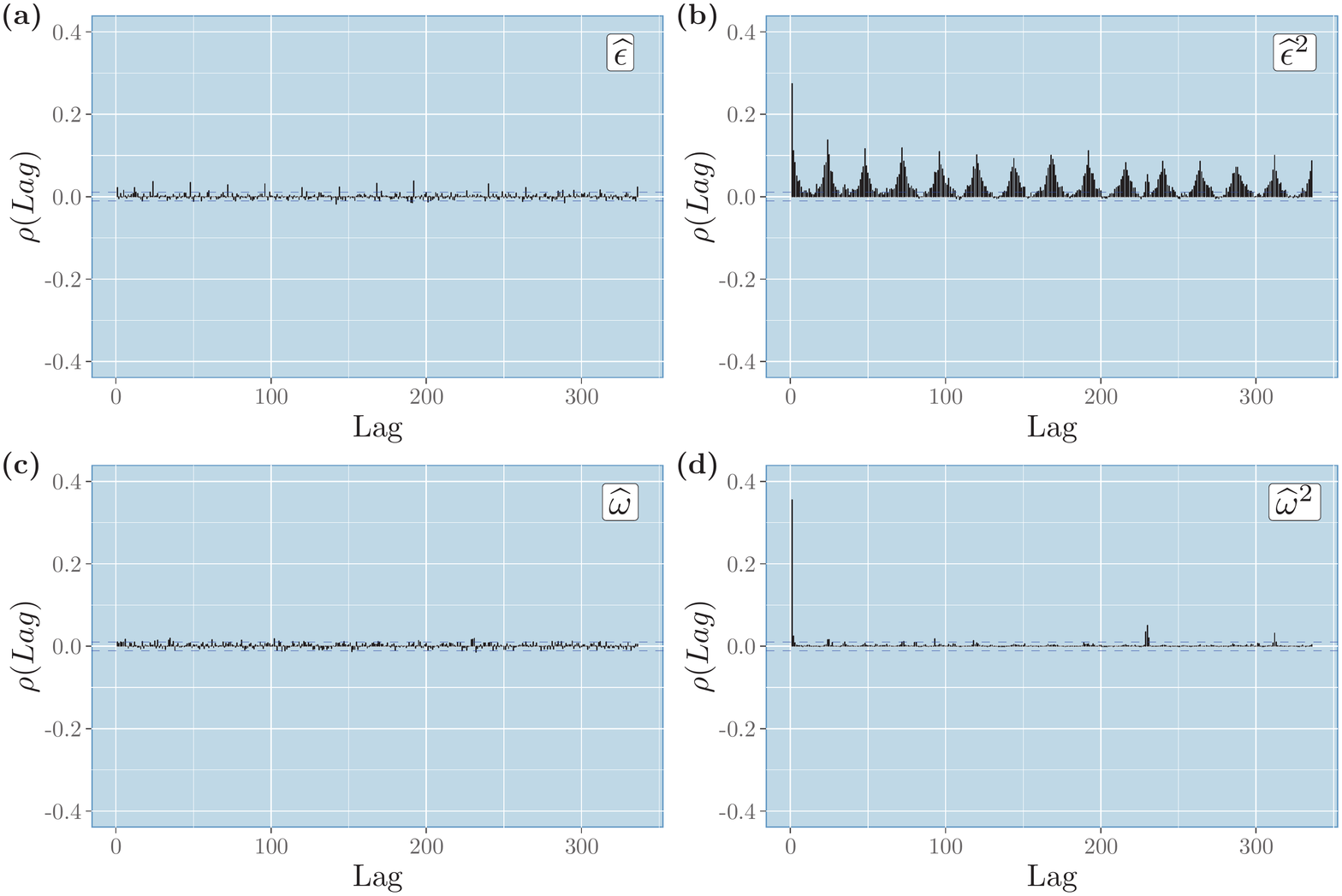}
\caption{Autocorrelation function of residuals $\widehat{\epsilon}\ \text{(a) and}\ \widehat{\epsilon}^2$ (b) of conditional mean estimation and autocorrelation function of residuals $\widehat{\omega}\ \text{(c) and}\ \widehat{\omega}^2$ (d) of conditional variance estimation.}
\label{fig_acf}
\end{figure}

\begin{figure}
\centering
\includegraphics[width=0.7\linewidth, angle =270 ]{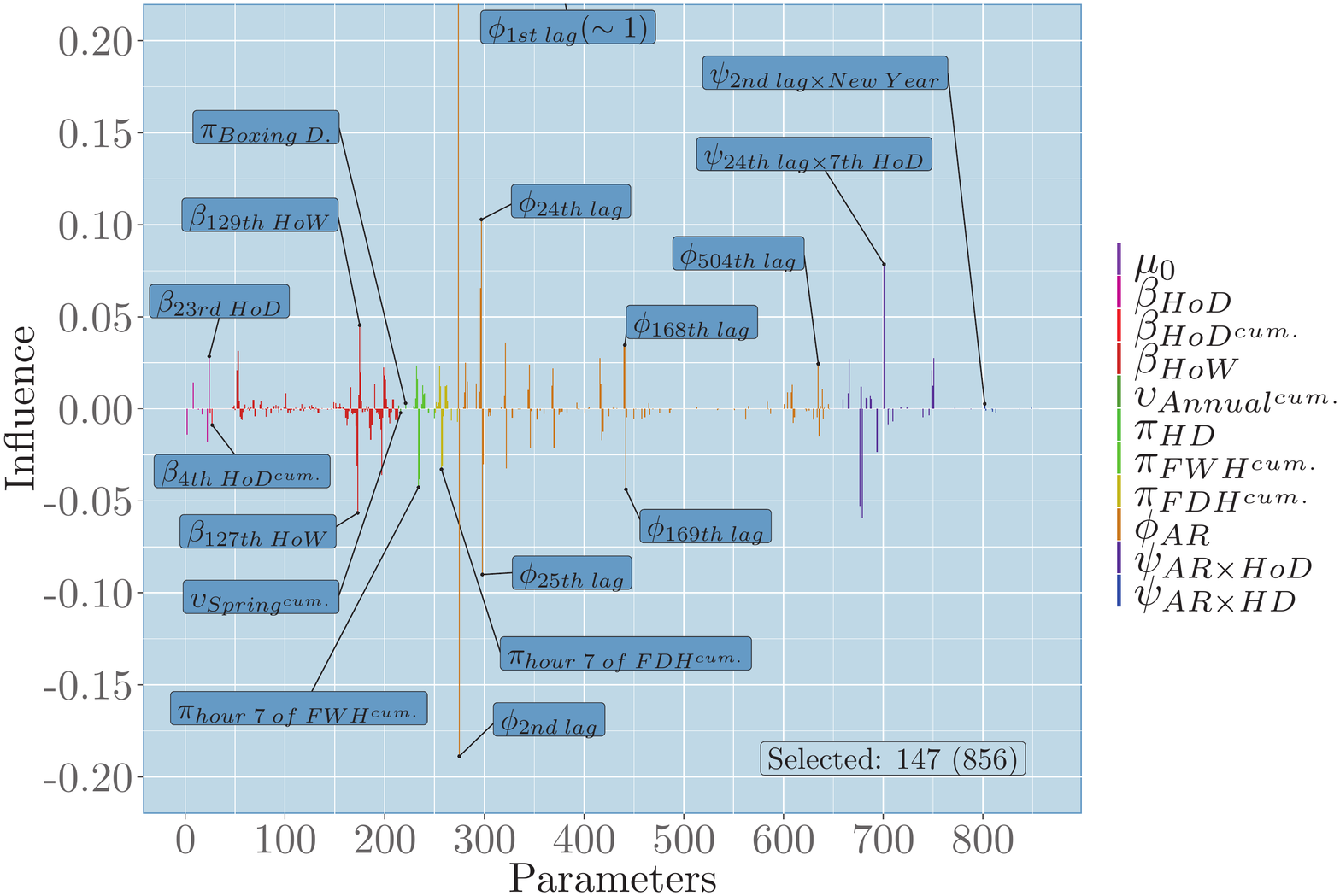}
\caption{Scaled parameters of conditional mean estimation.}
\label{fig:features}
\end{figure}

\begin{figure}
\centering
\includegraphics[width=0.7\linewidth, angle =270 ]{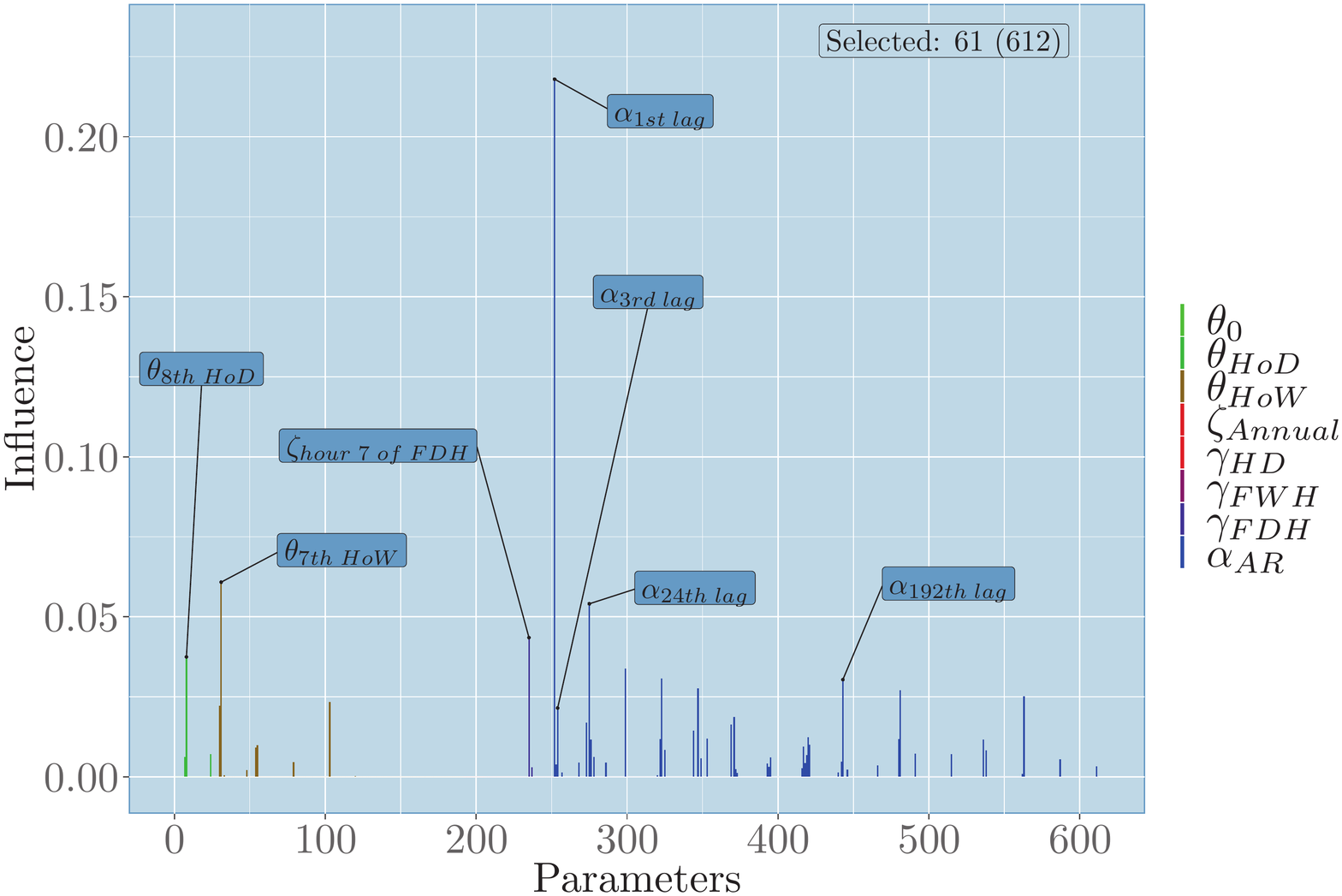}
\caption{Scaled parameters of conditional variance estimation.}
\label{fig:cvfeatures}
\end{figure}

\subsection{Results of validation period}
\begin{figure}
\centering
\includegraphics[width=1\linewidth]{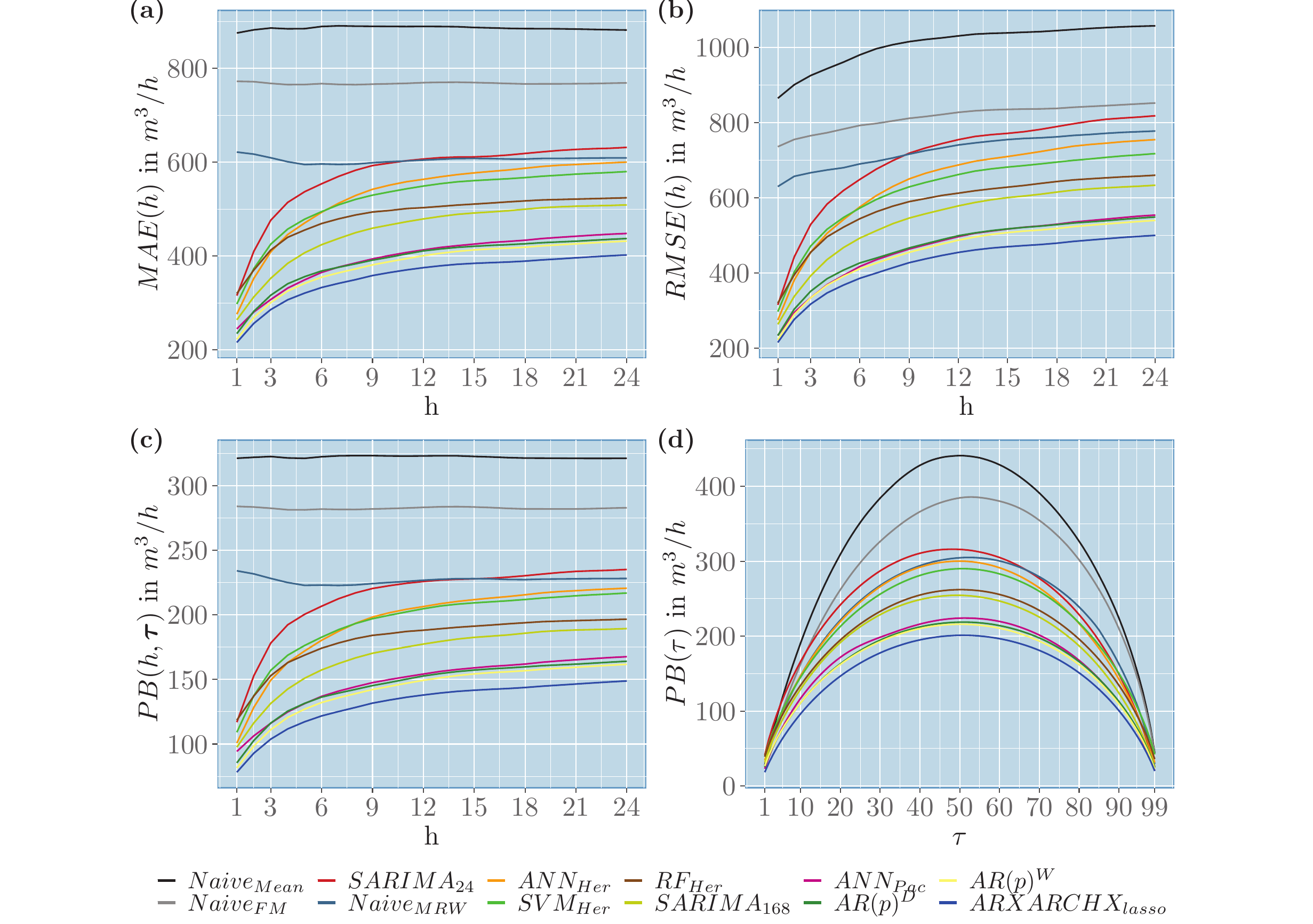}
\caption{Plot of MAE (a), RMSE (b), and PB (c) of one up to 24 hours and plot of PB (d) for the entire probability grid of all considered forecasting models for the validation period.}
\label{EC}
\end{figure}

The obtained scores in terms of the ES in Table \ref{V_Results}, show that the proposed forecasting model $ARXARCHX_{lasso}$ dominates all considered benchmarks significantly. For illustration, also the comonotone ($^{*}$), the countermonotone ($^{**}$), and the independent ($^{***}$) model simulations of the $ARXARCHX_{lasso}$ are presented. Here, it can be demonstrated, that neither the MAE, nor the RMSE, nor the NS or the PB are able to detect the manipulation in the dependency structure. However, although the ES is the determining measure in terms of evaluating a complete multidimensional distribution, pinpointing specific causes of a poor performance is rather difficult. Here, the PB, MAE, and RMSE allow a more thorough analyses. Focusing on the forecasting accuracy with expanding forecasting horizon from one up to 24 hours in Fig. \ref{EC} (a)-(c), it is observable, that with expansion of the horizon the forecasting error for all considered models increases notably. Here, the $ARXARCHX_{lasso}$ is constantly ranked the most accurate and reveals a moderate growth rate over time, compared to the benchmark models for all three criteria. Hence, the authors conclude that also for smaller forecasting horizons, the $ARXARCHX_{lasso}$ model is well suited. Focusing on Fig. \ref{EC} (d), the $PB(\tau)$ is plotted across the applied probability grid $\boldsymbol{\tau}$. Here, the $PB(\boldsymbol{\tau})$ of equation (\ref{PB}) has been computed without averaging across the quantile levels, so that explicitly the marginal properties in terms of the probabilities can be evaluated. It can be seen that the $ARXARCHX_{lasso}$ achieves the best score for each probability.
Hence, the superior performance of the $ARXARCHX_{lasso}$ can be verified at once in terms of the ES, but also for the mean and the marginal properties separately by considering the MAE, RMSE, and PB.\\
Focusing on the considered benchmark models, the forecasting accuracy of the  $SARIMA(0,1,4)(0,1,1)$ models has distinctively dropped compared to the calibration period. Here, beside a too short chosen calibration period, the authors suspect that the application of both an ordinary and a seasonal differencing operator might cause the rather poor performance. As most accurate competitors the proposed simple $AR(p)$ models can be highlighted and hence, the authors suggest them as easy implementable and fast computable benchmarks in the field of water demand forecasting.

\begin{table}[ht]
\caption{Forecasting results and performance improvements (Imp.) in \% relative to the $AR(p)^W$ model for each forecasting model within the validation period.}
\centering
\scalebox{0.7}{
\begin{tabular}{lrrrrrrrrrr}
 \hline
 Models & ES in & Imp. in & PB in & Imp. in & MAE in & Imp. in & RMSE in & Imp. in & NS in & Imp. in \\ 
  & $m^3/h$ &  \% &$m^3/h$ & \% & $m^3/h$ & \% & $m^3/h$ & \% & \% & \% \\ 
  \hline
$Naive_{Mean}$ & 3766.82 & -92.92 & 321.16 & -98.58 & 881.77 & -104.30 & 1056.75 & -95.75 & 69.65 & -23.62 \\ 
  $Naive_{FM}$ & 3173.94 & -62.56 & 282.89 & -74.93 & 769.01 & -78.18 & 851.77 & -57.78 & 77.98 & -14.49 \\ 
  $SARIMA(0,1,4)(0,1,1)_{24}$ & 2880.88 & -47.55 & 235.15 & -45.40 & 631.48 & -46.31 & 817.87 & -51.50 & 82.09 & -9.98 \\ 
  $Naive_{MRW}$ & 2775.66 & -42.16 & 228.15 & -41.08 & 609.12 & -41.13 & 777.25 & -43.98 & 81.14 & -11.02 \\ 
  $ANN_{Her}$ & 2693.59 & -37.96 & 220.64 & -36.43 & 600.14 & -39.05 & 754.47 & -39.76 & 83.06 & -8.91 \\ 
  $SVM_{Her}$ & 2602.74 & -33.30 & 216.90 & -34.12 & 579.92 & -34.37 & 717.37 & -32.89 & 84.83 & -6.98 \\ 
  $RF_{Her}$ & 2397.10 & -22.77 & 196.66 & -21.60 & 524.05 & -21.42 & 659.81 & -22.22 & 87.49 & -4.06 \\ 
  $SARIMA(0,1,4)(0,1,1)_{168}$ & 2272.23 & -16.37 & 189.38 & -17.10 & 508.89 & -17.91 & 633.23 & -17.30 & 87.38 & -4.18 \\ 
  $ANN_{Pac}$ & 2025.14 & -3.72 & 167.61 & -3.64 & 447.96 & -3.79 & 553.90 & -2.60 & 90.72 & -0.51 \\ 
  $ARXARCHX_{lasso}^{**}$ & 1990.54 & -1.95 & $\textbf{ 148.83 }$ & $\textbf{ 7.97 }$ & $\textbf{ 402.16 }$ & $\textbf{ 6.82 }$ & $\textbf{ 500.01 }$ & $\textbf{ 7.38 }$ & $\textbf{ 92.51 }$ & $\textbf{ 1.45 }$ \\ 
  $AR(p)^D$ & 1983.53 & -1.59 & 164.09 & -1.46 & 437.14 & -1.28 & 548.88 & -1.67 & 90.90 & -0.31 \\ 
  $AR(p)^W$ & 1952.51 & 0.00 & 161.72 & 0.00 & 431.60 & 0.00 & 539.84 & 0.00 & 91.19 & 0.00 \\ 
  $ARXARCHX_{lasso}^{*}$ & 1896.58 & 2.86 & $\textbf{ 148.83 }$ & $\textbf{ 7.97 }$ & $\textbf{ 402.16 }$ & $\textbf{ 6.82 }$ & $\textbf{ 500.01 }$ & $\textbf{ 7.38 }$ & $\textbf{ 92.51 }$ & $\textbf{ 1.45 }$ \\ 
  $ARXARCHX_{lasso}^{***}$ & 1808.32 & 7.39 & $\textbf{ 148.83 }$ & $\textbf{ 7.97 }$ & $\textbf{ 402.16 }$ & $\textbf{ 6.82 }$ & $\textbf{ 500.01 }$ & $\textbf{ 7.38 }$ & $\textbf{ 92.51 }$ & $\textbf{ 1.45 }$ \\ 
  $ARXARCHX_{lasso}$ & $\textbf{ 1780.13 }$ & $\textbf{ 8.83 }$ & $\textbf{ 148.83 }$ & $\textbf{ 7.97 }$ & $\textbf{ 402.16 }$ & $\textbf{ 6.82 }$ & $\textbf{ 500.01 }$ & $\textbf{ 7.38 }$ & $\textbf{ 92.51 }$ & $\textbf{ 1.45 }$ \\ 
   \hline
\multicolumn{11}l{\textbf{Hypothesis of the DM test, that the loss differential series between best and second best ranked model is zero, could be rejected}}\\
\multicolumn9l{\textbf{at the 0.001 significance level for each considered evaluation criterion.}}\\
 \hline
\multicolumn3l{$^*$ With comonotone model simulations.}\\
\multicolumn3l{$^{**}$ With countermonotone model simulations.}\\
\multicolumn3l{$^{***}$ With independent model simulations.}\\

\label{V_Results}
\end{tabular}}
\end{table}

\subsection{Diagnostic checking and model interpretation}
As the proposed forecasting model $ARXARCHX_{lasso}$ is constructed in a linear modelling framework, the authors are able to make inferences and hence can pinpoint the importance of each considered parameter for the water demand process. Moreover, by doing residual diagnostic shortcomings of the model can be outlined, so that further improvements can be suggested.\\
Considering the ACF of $\widehat{\epsilon}$ in Fig. \ref{fig_acf} (a), it is observable that the autocorrelation structure is sufficiently captured by the conditional mean estimation. Thus, without taking extra external information into account, no major forecasting improvements can be expected. Considering the ACF of $\widehat{\epsilon}^2$ in Fig. \ref{fig_acf} (b), it becomes obvious that volatility clusters exist. Therefore, the authors introduced equation (\ref{eq3}), to account for the time-varying variance. As illustrated in the ACF of the residuals of the conditional variance estimation $\widehat{\omega}$ in Fig. \ref{fig_acf} (c), the volatility clusters could be mostly captured. However, the strong first lag of the ACF of the squared residuals of the conditional variance estimation $\widehat{\omega}^2$ in Fig. \ref{fig_acf} (d), is suspicious and indicates that still further improvements are achievable. In this context, the authors would suggest introducing an absolute instead of a squared function for modelling the volatility cluster. This might be a promising task, as the residuals of the water demand process turned out to be heavy-tailed. Moreover, an improvement in accuracy might also be achievable by applying an iteratively reweighted adaptive lasso algorithm estimation procedure, as done by \citeN{Ziel.2016}, to better deal with the conditional heteroscedasticity in the high-dimensional setting.\\ 
Concentrating on the huge feature space, the linear forecasting framework allows for pinpointing the most influential features. As an automatic shrinkage and selection operator is applied, only a small fraction of the initially introduced 1,468 features is selected and only a sub fraction is considered as highly influential, as illustrated in Fig. \ref{fig:features} and Fig. \ref{fig:cvfeatures}. In this context, it has to be noted that the parameters are scaled, so that the influence of the features is illustrated in relation to each other and not their absolute effects.\\
For the conditional mean estimation in Fig. \ref{fig:features}, only 147 out of 856 features have been selected. Here, it is striking that all components contain influential features, what in turn might justify the consideration of the rather high-dimensional feature space to model the water demand process. By highlighting only the most important features, the authors can name hour 23 of the day and the change over time of hour 4 of the day; hour 127 and 129 of the week; hour 7 of the fixed weekday and fixed date holiday, respectively, as well as the autoregressive lags 1,2,23,24,25,167,168, and 169 and the interaction of lag 2 and hour 7. Overall, the autoregressive component can be identified as the most influential. This might also explain the superior performance of the proposed $AR(p)$ models. For instance, the $AR(p)^W$ model has selected in average an order of $p=233$, so that the previous 233 hours have been used. Considering the applied forecasting models in the literature, it can be said that most models have considered autoregressive effects, for example done by \citeN{Pacchin.2019}, \citeN{Anele.2017}, \citeN{Arandia.2016}, and \citeN{Chen.2018}. Moreover, deterministic patterns on the daily, weekly, and annual scale have also regularly been considered, for example done by \citeN{Alvisi.2017}, \citeN{Gagliardi.2017b}, and \citeN{Brentan.2017}. However, as mostly models with a low-dimensional feature space have been used the effects have not been modelled in a high resolution. Here, the authors conclude that the difference in forecasting accuracy can be explained by the difference in the amount of initially considered information and the model itself to deal with the provided information. Besides, it might be worth noting that the number of parameters does not necessarily corresponds to the number of features. Hence, although the $ARXARCHX_{lasso}$ seems to be rather complex, the number of active parameters used in the final model is only moderate compared to the number of parameters of the introduced machine learning based models, as shown in Table \ref{C_Results}.\\
For the conditional variance estimation in Fig. \ref{fig:cvfeatures}, only 61 out of 612 features are selected. Here, only the hour of the day, the hour of the week, the fixed date holiday and the autoregressive component are considered as influential. By highlighting the most important features of each component, the authors can name hour 8 of the day; hour 7 of the week; hour 7 of a fixed weekday holiday and the autoregressive lags 1,24,48, and 192. By examining the existing literature, only a few modelling approaches have addressed the time-varying variance structure, for example done by \citeN{Caiado.2010} and \citeN{Hutton.2015}.

\begin{figure}[htbp]
    \centering
    \begin{subfigure}[b]{\textwidth}
        \centering
        \includegraphics[width=.32\textwidth]{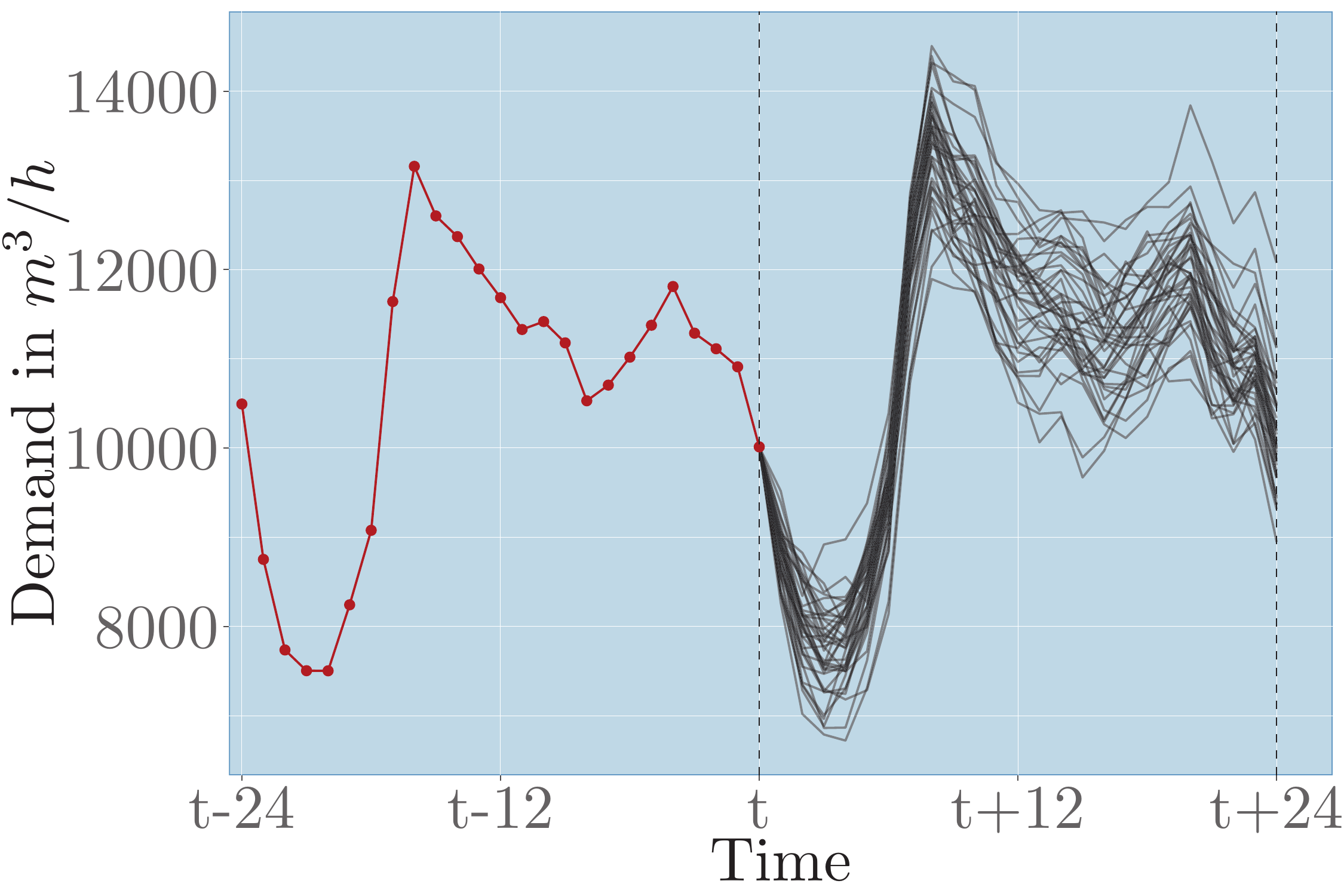}\hfill
        \includegraphics[width=.34\textwidth]{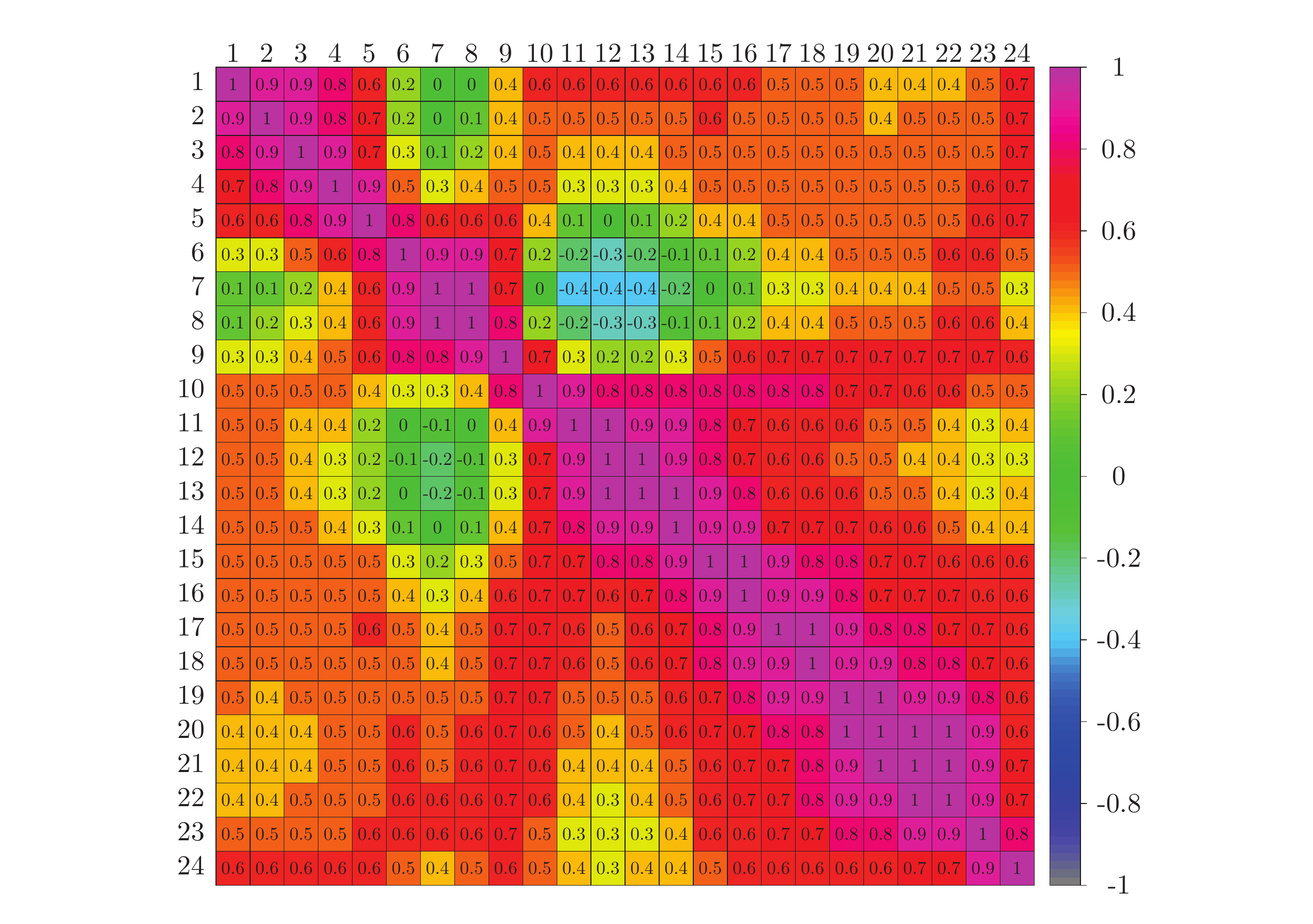}\hfill
        \includegraphics[width=.32\textwidth]{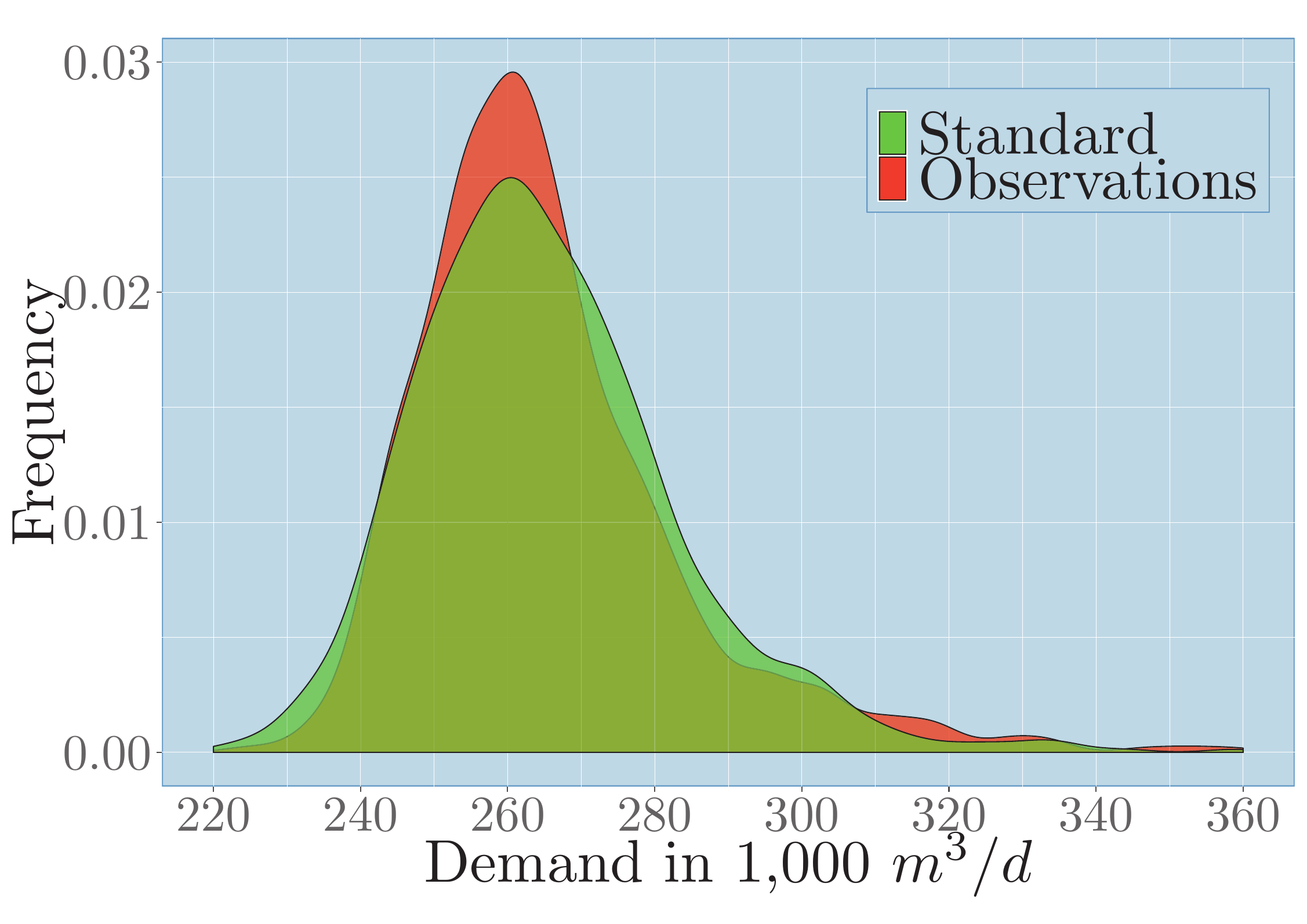}
    \caption{Standard model simulations.}
    \end{subfigure}
        \begin{subfigure}[b]{\textwidth}
        \centering
        \includegraphics[width=.32\textwidth]{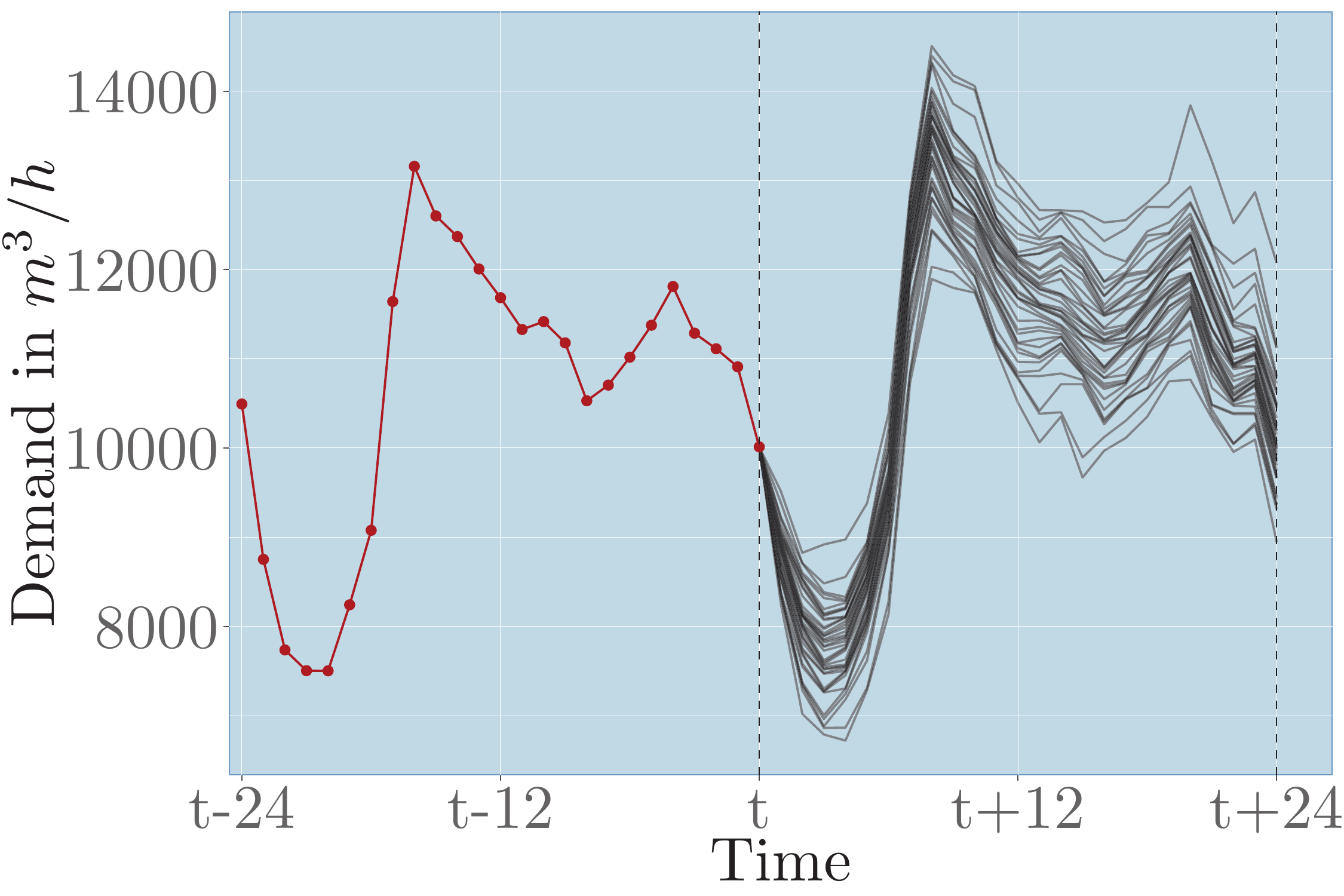}\hfill
        \includegraphics[width=.34\textwidth]{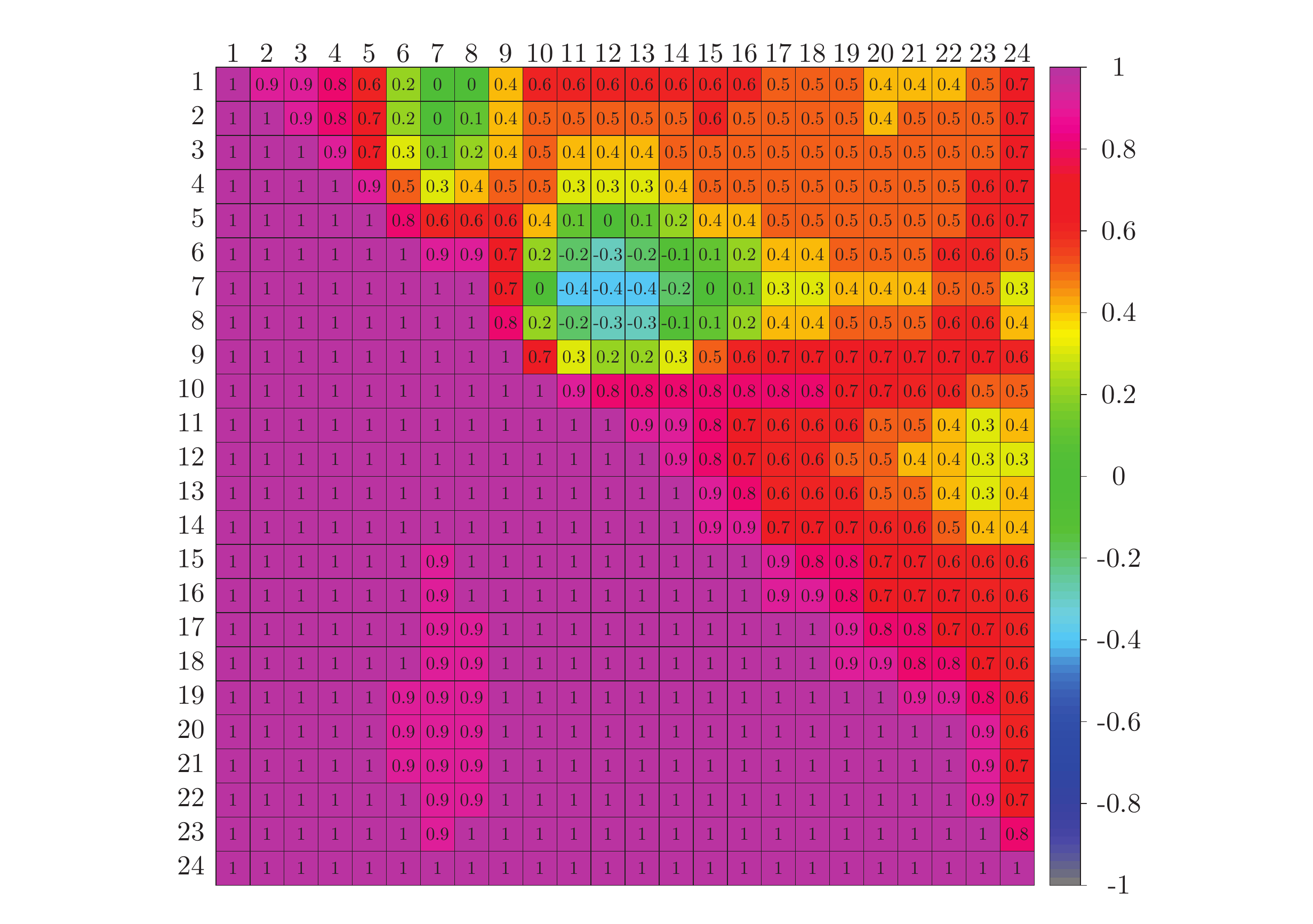}\hfill
        \includegraphics[width=.32\textwidth]{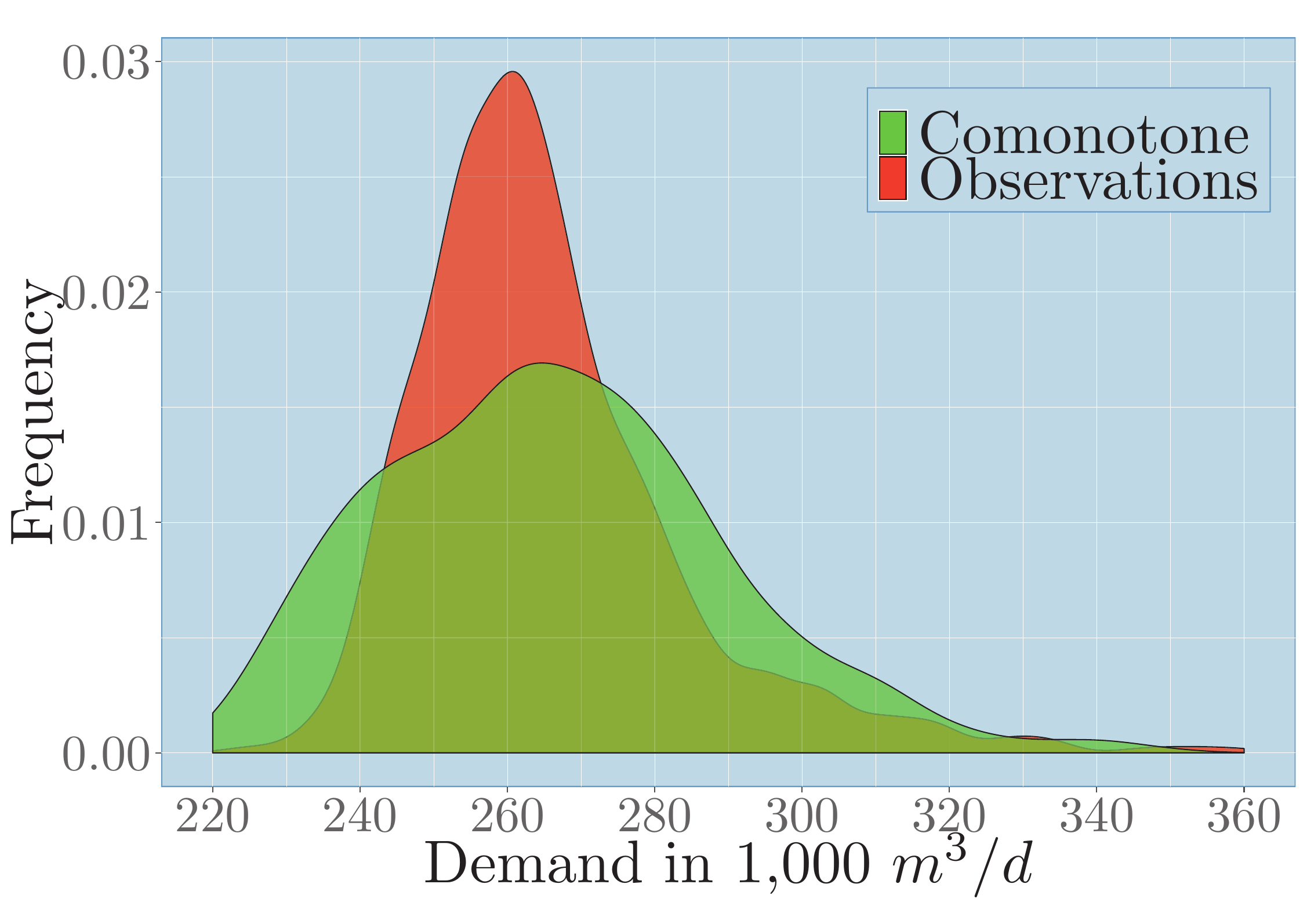}
    \caption{Comonotone model simulations.}
    \end{subfigure}
         \begin{subfigure}[b]{\textwidth}
        \centering
        \includegraphics[width=.32\textwidth]{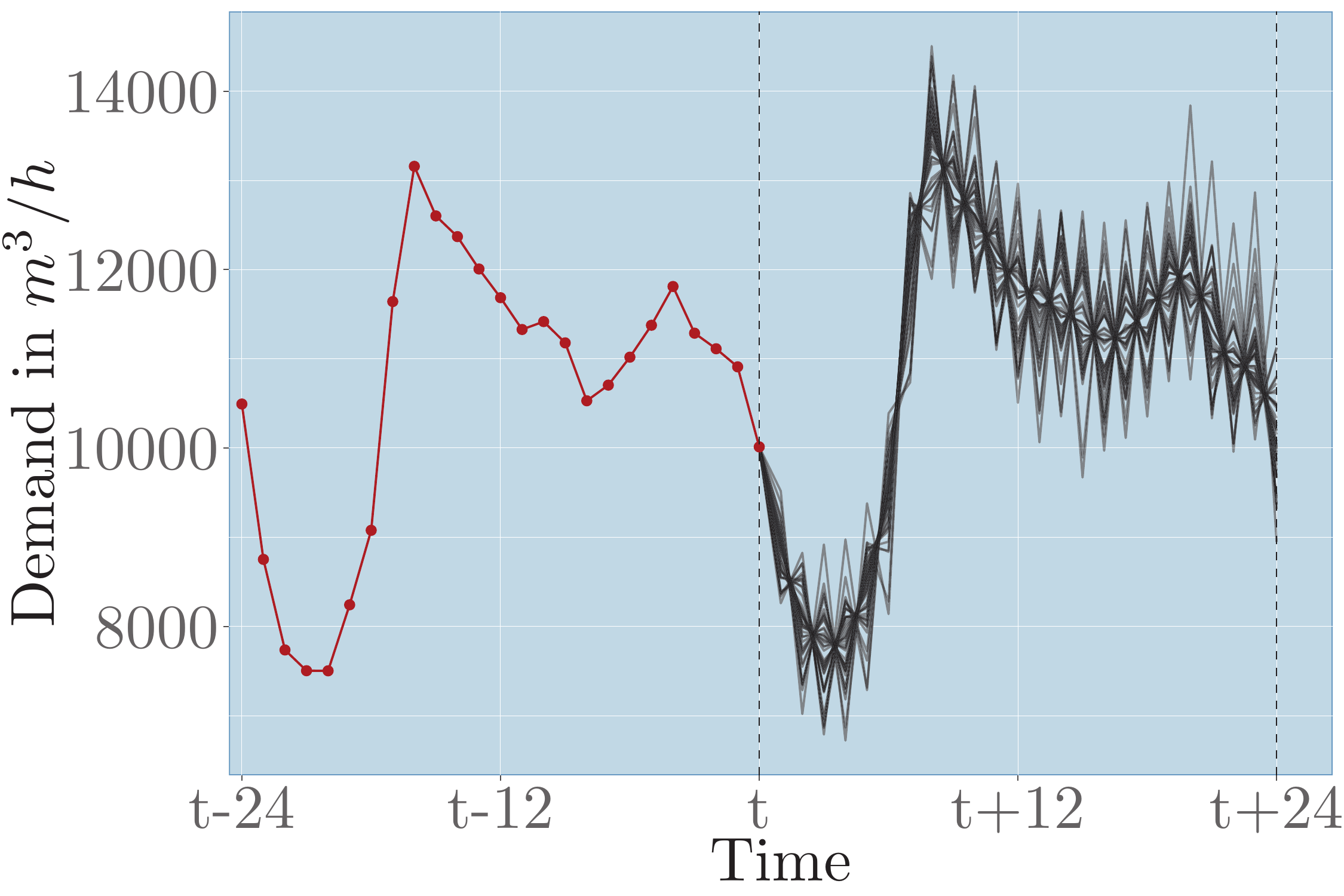}\hfill
        \includegraphics[width=.34\textwidth]{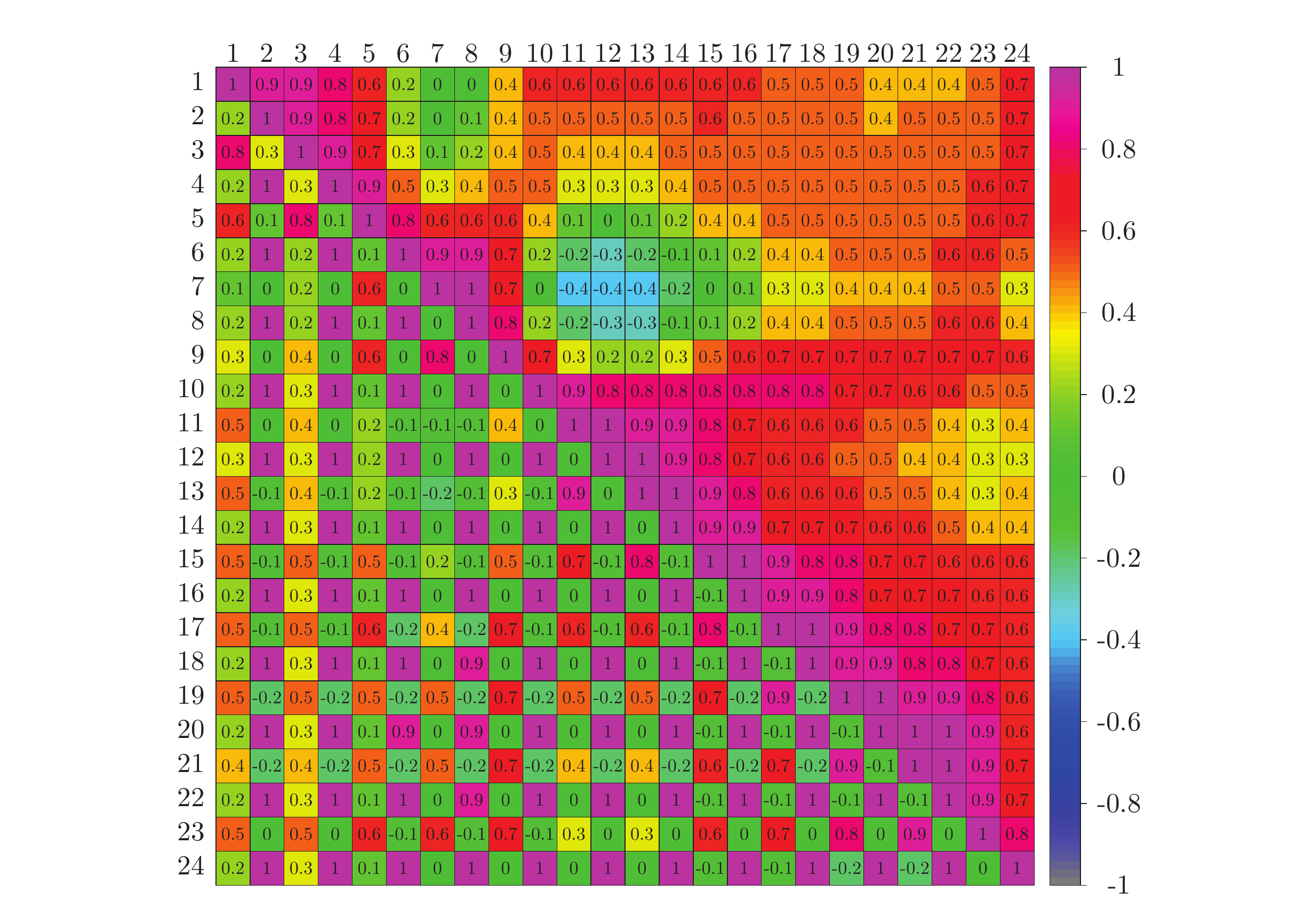}\hfill
        \includegraphics[width=.32\textwidth]{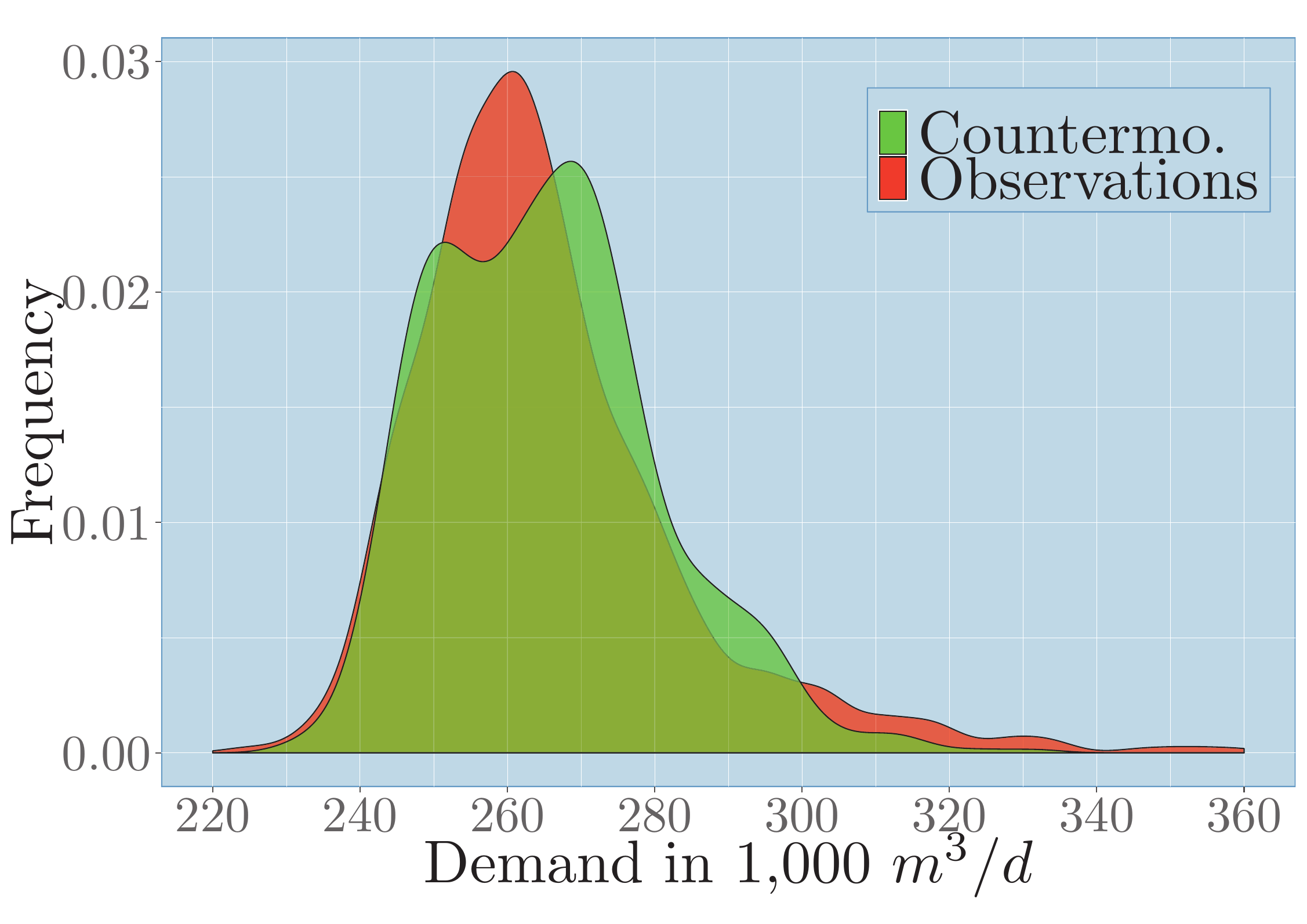}
    \caption{Countermonotone model simulations.}
    \end{subfigure}
        \begin{subfigure}[b]{\textwidth}
        \centering
        \includegraphics[width=.32\textwidth]{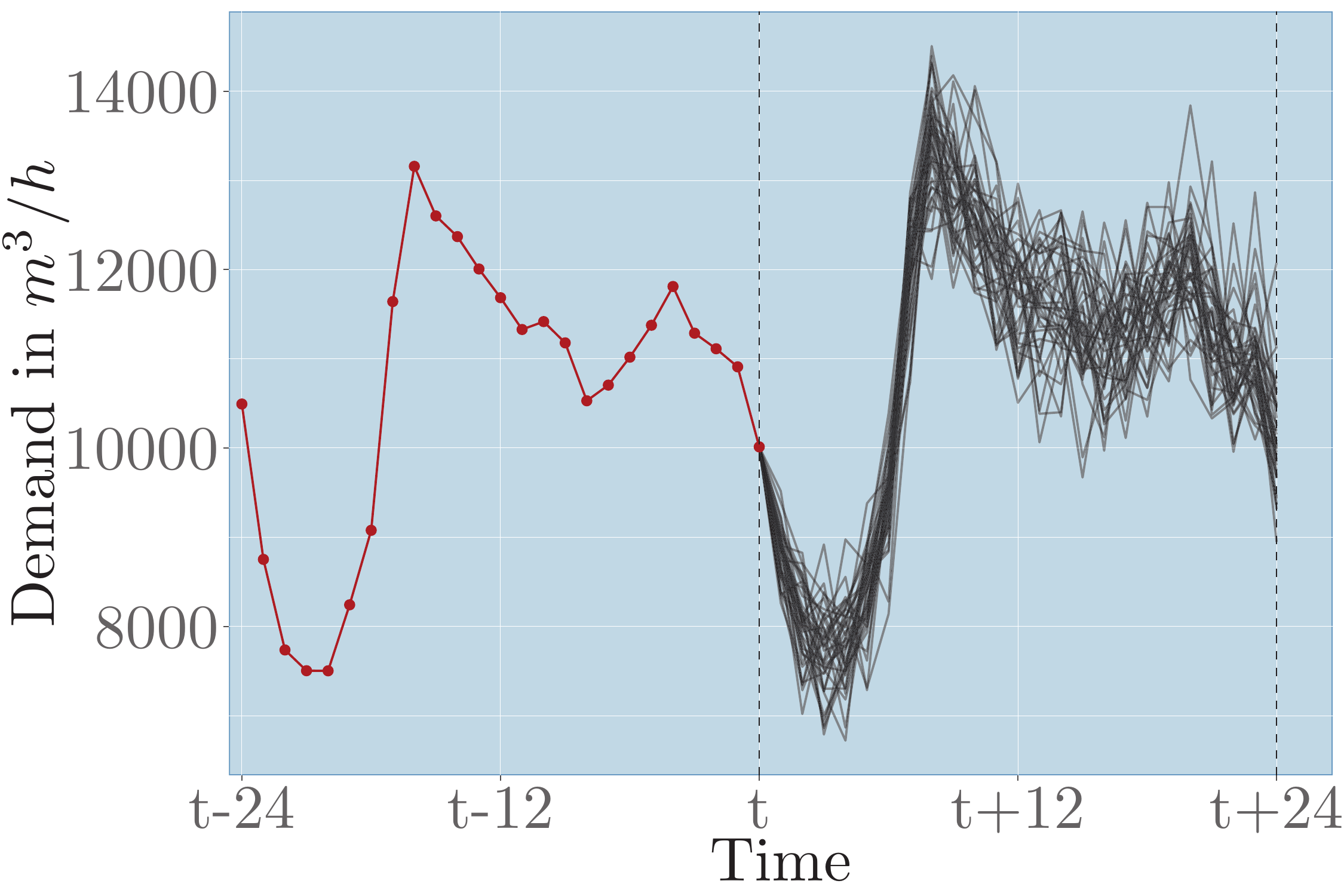}\hfill
        \includegraphics[width=.34\textwidth]{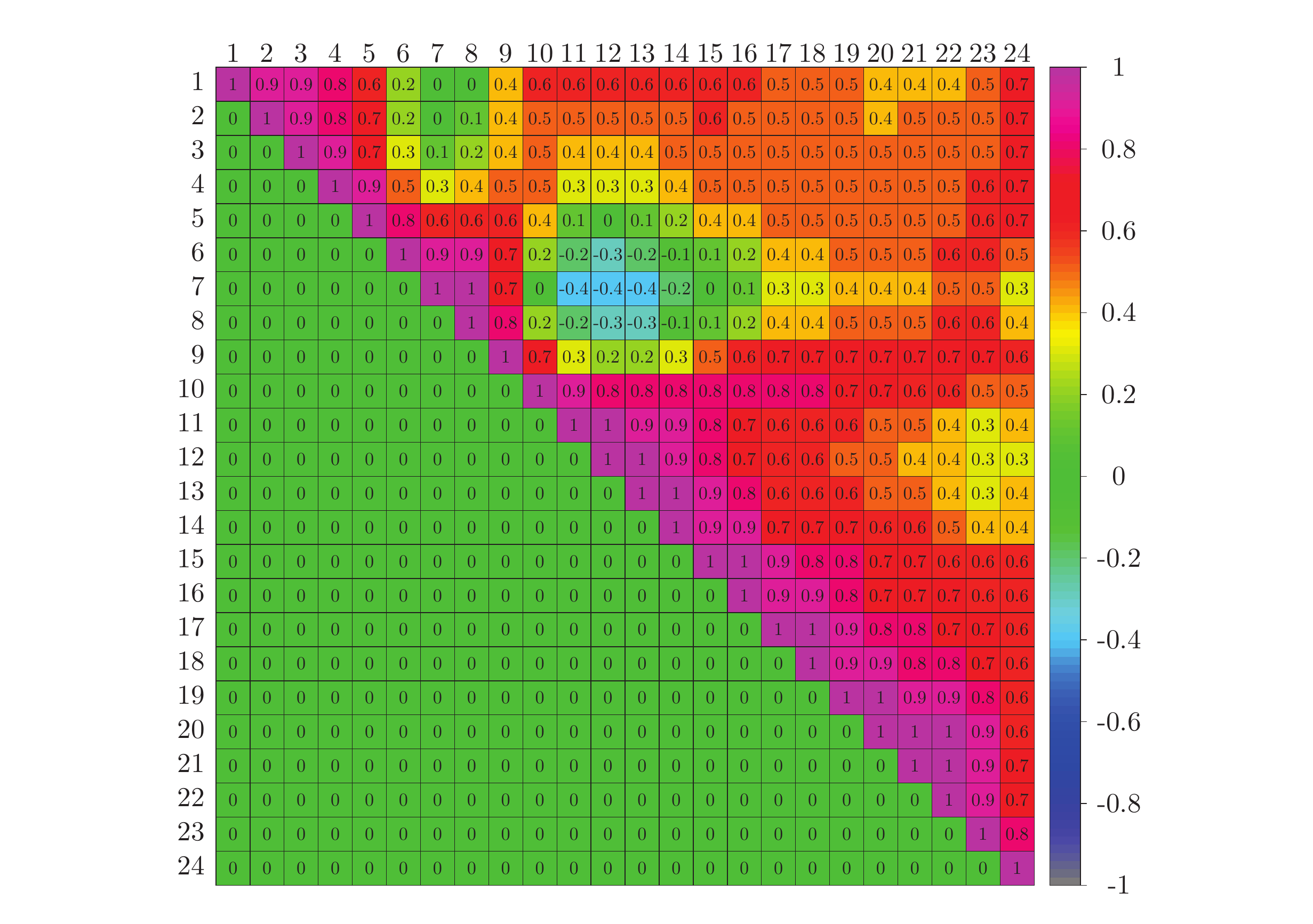}\hfill
        \includegraphics[width=.32\textwidth]{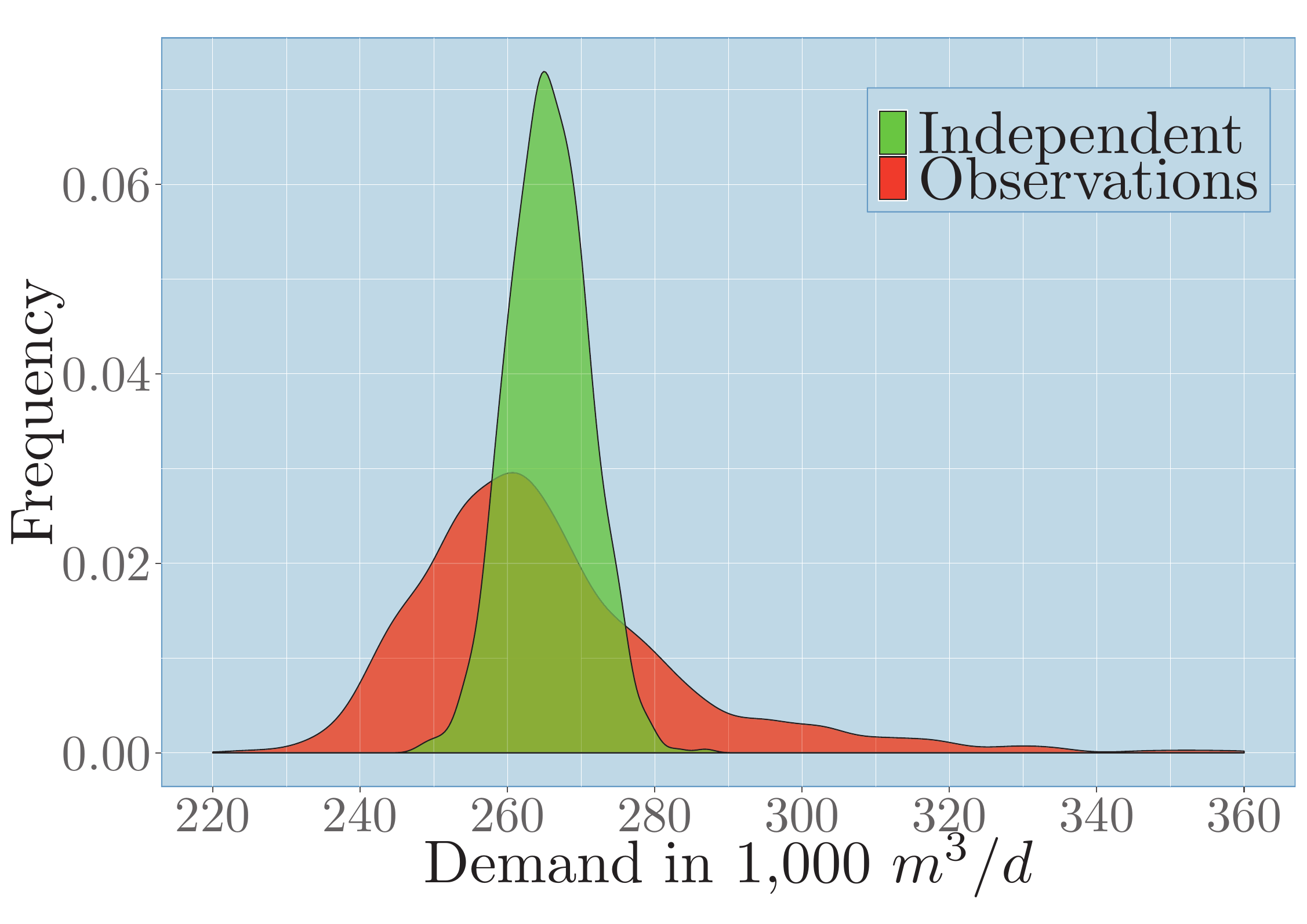}
    \caption{Independent model simulations.}
    \end{subfigure}
    \caption{Illustration of model simulations (left), simulated $24\times24$ correlation matrix (lower triangular) in comparison to true correlation matrix (upper triangular) (middle) and density of cumulative sample path of simulated and true water demand (right).}
    \label{fig_CD}
\end{figure}

\subsection{Discussion on practicality of complete probabilistic multi-step-ahead forecasts}
\begin{figure}
\centering
\includegraphics[width=1\linewidth]{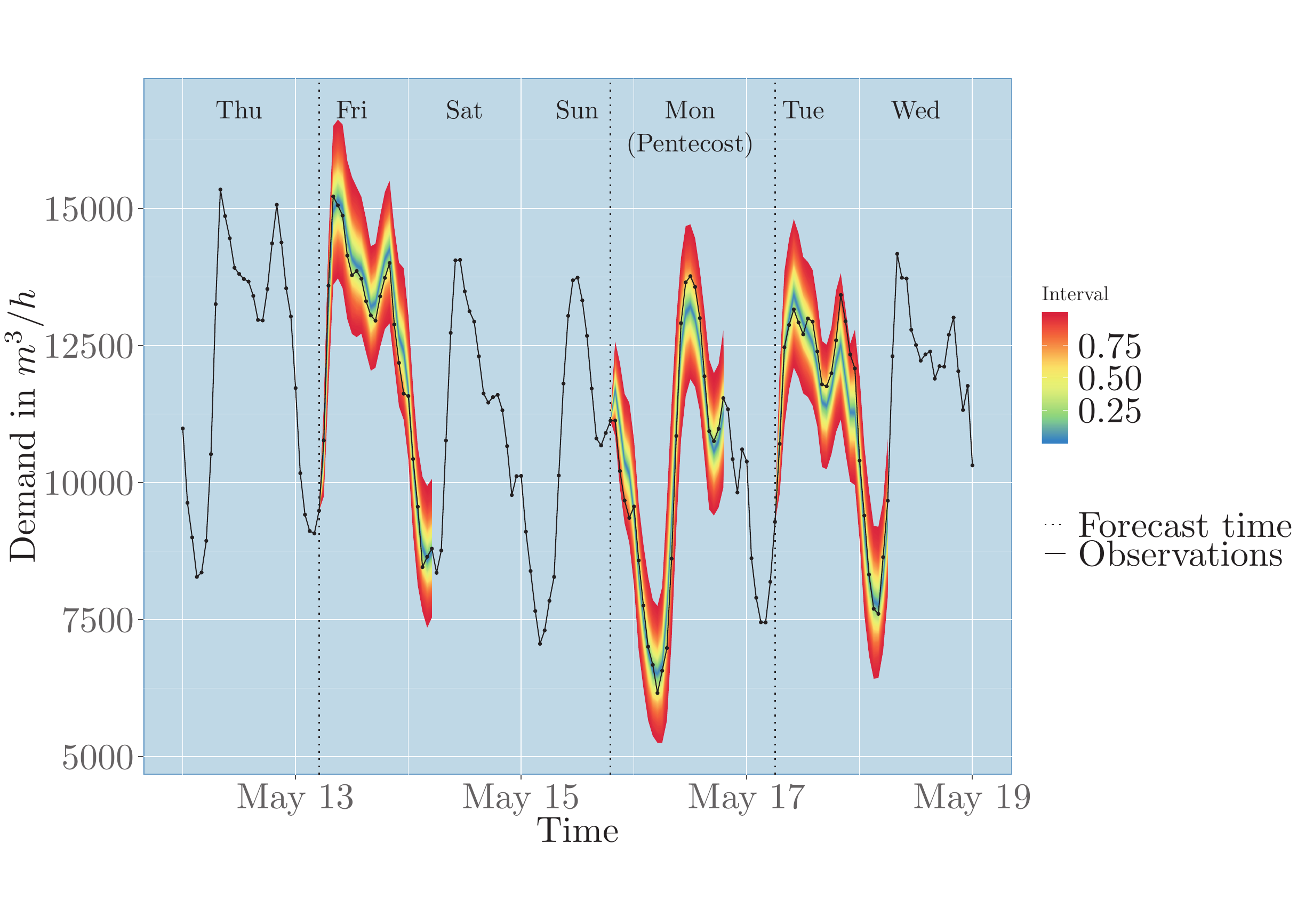}
\caption{Plot of three complete probabilistic multi-step-ahead forecasts within a week in 2016. The obtained ES in $m^3/h$ from left to right: $1181.82$; $1563.34$; and $1555.53$.}
\label{fig_PM}
\end{figure}

Besides the mean and the marginal properties is the dependency structure of probabilistic multi-step-ahead forecasts of considerable relevance for practical applications, for example, with respect to time-dependent optimization problems in the field of water planing and management. For illustration, the proposed $ARXARCHX_{lasso}$ model and the manipulated versions of the $ARXARCHX_{lasso}$ model of Table \ref{V_Results} are considered. They are identical in the mean and the marginal properties but differ in the simulated dependency structure, as highlighted in Table \ref{V_Results} and displayed in Fig. \ref{fig_CD}. Focusing on the accuracy of the simulated dependency structures, the $24\times24$ correlation matrix of the corresponding model simulation (lower triangular) compared with the correlation matrix of the true water demand (upper triangular) is displayed in the middle column of Fig. \ref{fig_CD}. It is obvious, that only the standard model simulation (a) covers the true dependency structure of the water demand process reasonably. To verify the practicality, the water storage optimization as an actual operative management is considered. It forms the foundation to better balance demand peaks, to increase the security of supply, and to better schedule the pumping arrangements to take advantage of the electricity price structure. Here, decision makers are especially interested in the expected aggregated or cumulative demand, so that a statement can be made about the probability with which a water storage capacity can guarantee the supply over a certain period of time.
The impact of the dependency structure on the cumulative water demand is illustrated on the right column of Fig. \ref{fig_CD}. Here, the density of the cumulative sample paths of each model simulation is compared with the true density of the cumulative water demand process.\\
Moreover, assuming a hypothetical water storage tank with a capacity of 290,000 m³ and a considered time period of 24 hours, the actual expected probability for each model simulation is computed and compared with the true probability, as shown in Table \ref{probs}. Here, the true cumulative water demand exceeds with a probability of 0.0935 the storage capacity within the corresponding 24 hours. The $ARXARCHX_{lasso}$ with a standard model simulation predicts a probability of 0.0934, the comonotone model simulation a probability of 0.1524, the $ARXARCHX_{lasso}$ with a countermonotone model simulation a probability of 0.0782, and the $ARXARCHX_{lasso}$ with an independent model simulation a probability of 0.0000. Hence, although all model simulations are simulated with identical mean and marginal properties, their expected cumulative water demand distributions differ distinctively and only the standard model simulation (a) of the proposed $ARXARCHX_{lasso}$ provides a reasonable forecast.\\
Finally, to get an impression of the forecasting performance of the $ARXARCHX_{lasso}$ model, in Fig. \ref{fig_PM}, three forecasts of the validation period are presented. Beside two regular days (on the right and left side) also one fixed weekday holiday is illustrated. With respect to the presented holiday, it is striking that especially the early morning hours are well captured, as the demand distinctively drops below the usual demand on a weekday and even below the demand of the preceding Saturday and Sunday. Concerning the overall performance of the proposed model, the authors determined that even on average the forecasting performance is satisfactory; however special days such as specific holidays and other rarely occurring events are extremely difficult to capture. This is also valid for some periods during the summer. Here, the extension of the calibration period with a focus on rarely occurring events and holidays or the introduction of weather forecasts might lead to improvements. Nevertheless, on normal days the proposed model performs very well and is therefore well suited for a large part of the year.

\begin{table}
\caption{Probabilities that the true and the simulated cumulative water demands exceed the water storage capacity of 290,000 $m^3$ within a time period of 24 h.}
\centering
\scalebox{0.8}{
\begin{tabular}{lccccc }
\hline
  & Observations & $ARXARCHX_{lasso}$ & $ARXARCHX_{lasso}^{*}$ & $ARXARCHX_{lasso}^{**}$ & $ARXARCHX_{lasso}^{***}$  \\
  \addlinespace[0.1cm]
\hline
   Probabilities  & 0.0935 & 0.0934 & 0.1524 & 0.0782 & 0.0000 \\
\hline
\multicolumn3l{$^*$ With comonotone model simulations.}\\
\multicolumn3l{$^{**}$ With countermonotone model simulations.}\\
\multicolumn3l{$^{***}$ With independent model simulations.}\\
\label{probs}
\end{tabular}}
\end{table}

\section{Summary and Conclusion}
This paper proposed a forecasting model with a high-dimensional feature space in a linear framework to capture the complex structure of the water demand process. By applying the least absolute shrinkage and selection operator (lasso), the model could be automatically tuned, so that a parsimonious, simple interpretable and fast computable forecasting model could be obtained. The model clearly outperformed existing benchmarks from the water demand forecasting literature. Here, the authors concluded that the complex water demand process can be more accurate modelled by automatically tuned linear models with a high-dimensional feature space than by applying rather computational complex non-linear models with a low-dimensional feature space.\\
Moreover, an appropriate multi-step-ahead forecasting framework to issue a complete multivariate probabilistic forecasting distribution, which is able to account also for path-dependencies, has been introduced and their practicality in the field of water storage management has been verified. Furthermore, the need for more sophisticated evaluation measures has been outlined and the ES as a strictly proper scoring rule has been proposed. The ES allows for penalizing errors in the mean, the marginal properties and the correlation structure of the corresponding forecast. 
Nevertheless, there is still much research to be done. Focusing on the proposed forecasting model, it might be worthwhile to include forecasts of weather inputs or further external deterministic processes such as the announced water withdrawal in advance of bulk customers. Concerning the already considered features, the extension of the time-varying autoregressive component and applying more advanced modelling techniques for holidays and the time-varying variance might be promising avenues to explore.

%
\section{Data Availability statement}
Some or all data, models, or code generated or used during the study are proprietary or confidential in nature and may only be provided with restrictions (e.g. anonymized data).

\appendix
\begin{appendices}
\numberwithin{equation}{section}
\makeatletter 
\newcommand{\section@cntformat}{Appendix \thesection:\ }
\makeatother

\section{}
\label{appendix:A1}
The B-spline basis function of degree $H$ can be modelled by a simple B-spline basis function $\widetilde{B}$, as shown by \citeN{Ziel.2016c}. The $\widetilde{B}$ is defined by the degree $H$ and a set of knots $\mathcal{K}$, whereby $\mathcal{K}$ contains $H+1$ knots $\{k_0,...,k_{H+1}\}$, with $k_h < k_{h+1}$. In accordance with \citeN{Boor.2001}, the recurrence relation is defined as:

\begin{equation}
\begin{split}
& \widetilde{B}(t;\{k_0,...,k_{H+1}\},H) = \\ & \frac{t-k_0}{k_H-k_0}\widetilde{B}(t;\{k_0,...,k_H\},H-1) + \\
& \frac{t-k_1}{k_{H+1}-k_1}\widetilde{B}(t;\{k_1,...,k_{H+1}\},H-1)
\label{rr}
\end{split}
\end{equation}
with initialization 
\[
\widetilde{B}(t;\{k_l,k_{l+1}\},0)= 
\begin{cases}
1 \quad,t\in [k_l,k_{l+1})\\
0 \quad, \text{otherwise}.
\end{cases}
\]
The set of knots $\mathcal{K}(T,H)$ is equidistant with center $T$. Hence, $k_0=T-h\frac{D+1}{2},k_{D+1}=T+h\frac{D+1}{2}$ and as an odd degree $D$ is selected, $k_{\frac{D+1}{2}}=T$ is obtained, whereby $h$ denotes the distance between the knots. Note, $H$ and $h$ define the knots $\mathcal{K}$ uniquely.\\
To obtain a periodic basis function $\widetilde{B}(t;\mathcal{K},H)$, a seasonality $S$ is required. It is suitable to choose $h$ such that $S$ is an integer multiple of $h$, which itself is at least $H+1$ to ensure a partition of the unity. The initial periodic basis function can be defined as

\begin{equation}
\widetilde{B}_1^*(t;\mathcal{K},H)=\sum_{k\in\mathbb{Z}}{ \widetilde{B}(t-kS;\mathcal{K},H)}
\label{ipb}
\end{equation}. 

In the corresponding setting, the data has three seasons, a diurnal, a weekly and an annual one. However, only the latter is modelled. As the considered data has one observation per hour, the annual season corresponds to $S_{\text{annual}} = 365.24x24=8765.76$. Note that an average year lasts 365.242375 days, which is approximated by the leap year system every four years. Hence, a year is approximated by 365.24 days. By using the initial periodic basis function $\widetilde{B}_1^*$, the full periodic basis can be defined by $\widetilde{B}_j^*(t;\mathcal{K},H) = \widetilde{B}_{j-1}^*(t-h;\mathcal{K},H)$. In conclusion, the basis $\mathcal{B}=\{\widetilde{B}_1^*,...,\widetilde{B}_{N_\mathcal{B}}^*\}$ has a total of $N_{\mathcal{B}} = S/h$ basis functions. In the corresponding setting, the authors choose $h_{\text{annual}} = 4$.\\
The basis functions $\widetilde{B}_l^*$ are suitable to capture seasonal changes of inputs. However, they model the absolute impact over time due to the introduced structure of the basis functions. In practice the consideration of changes over time are likewise promising, especially if an automatic shrinkage and selection algorithm is used for estimation. The changes in the parameters over time can be easily constructed by accumulating the basis function $\widetilde{B}_l^*$ within $l$. Hence, the authors define

\begin{equation}
\widetilde{B}_l^{*,\text{cum.}}=\widetilde{B}_{l-1}^{*,\text{cum.}}+\widetilde{B}_l^{*}
\label{ipbcum}
\end{equation}

for $l>1$ with $\widetilde{B}_l^{*,\text{cum.}}=\widetilde{B}_{1}^{*}$.
In the corresponding setting, the cumulative basis functions are used for the conditional mean model in equation (\ref{mu}) and non-cumulative basis functions are applied for the conditional variance model in equation (\ref{a_0}) due to the non-negativity parameter constraints.

\section{}
\label{appendix:B}
\begin{lstlisting}[language=R]
far<- function(y, H, M, p, S=168){
  # y = data, H = forecasting horizon,
  # M = number of simulations, p= max. order
  how <- matrix(, S, ceiling(length(y)/S))
  how[1:length(y)]<- y
  HOW<- rep_len(apply(how, 1, mean, na.rm=TRUE),
                length.out=length(y)+H)
  X<-y-HOW[1:length(y)]
  ar.model<- ar(X,
                order.max=p, method="yw") # estimation
  
  lagm<- ar.model$order
  get.lagged<- function(lag, Z){
    c( rep(NA, lag),  Z[(1+lag):(length(Z)) -lag ]  )
  }
  yLAG<- sapply(1:lagm, get.lagged, Z=c(X,NA)) 
  Xf<- array(, dim=c(lagm+H,lagm))
  Xf[1:lagm,]<- tail(yLAG , lagm) 
  Xfens<- array(Xf, dim=c( dim(Xf),M)) - ar.model$x.mean 
  yfens<- array( c(tail(X, lagm), rep.int(NA,H)),
                 dim=c( lagm+H,M))
  EPSfens<- array(sample(ar.model$res[!is.na(ar.model$res)],
                         size=H*M, replace=TRUE), dim=c(H, M))
  for(h in 1:H){ ## forecasting
    yfens[lagm+h,]<-
      as.numeric(ar.model$ar) %*% Xfens[lagm-1+h,,] +
      ar.model$x.mean +  EPSfens[h,]
    if(h<H)    
      Xfens[lagm+h,,]<-
        yfens[0:lagm+h,][lagm:1+1,]-ar.model$x.mean
  }
  tail(yfens,H) + tail(HOW,H) #return forecast
}
## illustration 
y<- arima.sim(n=1000, 
              model=list(ar=c(.7,-.1,0,0,0,0,0,0,0,0,0,.7,-.7) ))
ensemble  = far(y, H=24, M=10, p=50)
ts.plot( rbind(tail(y,1),ensemble), col=rainbow(10, alpha=.7) )



\end{lstlisting}

\end{appendices}

\bibliography{bibfile}

\end{document}